\documentclass[12pt]{article}
\pdfoutput=1
\usepackage[table]{xcolor}
\usepackage{cite}
\usepackage{avant}
\usepackage{amssymb}
\usepackage{multirow,booktabs}
\usepackage{caption}
\usepackage{subcaption}
\usepackage{pifont}\usepackage{amsmath}
\usepackage[mathscr]{eucal}
\usepackage[all]{xy}
\usepackage{listing}
\usepackage{pdfpages}
\usepackage{graphicx}
\xyoption{arc}
\xyoption{curve}
\numberwithin{equation}{section}
		\newfont{\spfnt}{punk12}
		\newcommand\email[4]{#1@#2.#3.#4}
		\def\F{\mathcal{F}}

		\renewcommand\hat{\widehat}

		\newcommand\be{\begin{equation}}
		\newcommand\ee{\end{equation}}
		
		\def\pa{\partial}

		\newfont\sheafnt{rsfs10}

\begin{document}
			\title{Holographic Fermi surfaces in charge density wave from D2-D8}
			\author{
				Subir Mukhopadhyay\thanks{\email{subirkm}{gmail}{com}{}} 
				 \, and 
				Nishal Rai\thanks{\email{nishalrai10}{gmail}{com}{}}		
			}
			\date{
			$^*$ {\small Department of Physics, Sikkim University, 6th Mile, Gangtok 737102.}\\
			$^\dagger$ {\small Departamento de F\'isica At\'omica, Molecular y Nuclear, Universidad de Granada, 18071-Granada, Spain.\\and\\ $^\dagger$Department of Physics, SRM University Sikkim, 5th Mile, Tadong, Gangtok 737102.}
			 }
			\maketitle
			%\vfil
			\begin{abstract}
				\small
				\noindent
				 D2-D8 model admits a numerical solution that corresponds to a charge density wave and a spin density wave. Considering that as the background, we numerically solve the Dirac equation for probe fermions. From the solution, we obtain the Green's function and study the behaviour of the spectral density. We begin with generic fermions and have studied the formation of the Fermi surface and where it develops a gap. In addition, we have incorporated an ionic lattice and study its effect on the Fermi surface. Then we analysed the worldvolume fermions. In this particular model we do not find Fermi surface for the dual operators.
			\end{abstract}
			\thispagestyle{empty}
			\clearpage

%%%	
		\section{Introduction}
		
There are certain strongly correlated electron systems, such as, strange metals in high T$_c$ superconductors and heavy fermion systems \cite{verma,anderson,Gegen}, which admit Fermi surfaces but the excitations are not long-lived and thus live outside the regime of the Fermi liquid theory. Techniques of holography \cite{Maldacena:1997re,Gubser:1998bc,Witten:1998qj,review} have been proved to be quite successful in unravelling various features of such non-Fermi liquids. Especially measurements of Angle Resolved Photoemission Spectroscopy (ARPES) or Scanning Tunneling Microscopy (STM) can be compared with holographic spectral functions leading to a test of holographic applications. Substantial studies of holographic Fermi surfaces have been conducted both in bottom-up \cite{Lee:2008xf, hening, mueck, Cubrovic:2009ye, Liu:2009dm, Faulkner:2009wj, Edalati:2010ge, Edalati:2010ww} and top-down \cite{Ammon:2010pg, Jensen:2011su, gauntlett, baillard, gauntlett1, berkooz, berkooz1, berkooz2, DeWolfe:2011aa, DeWolfe:2012uv, DeWolfe:2014ifa, Cosnier-Horeau:2014qya, self, self1} approaches (see \cite{iqbal} for a review) over the recent past. These works mostly deal with the homogeneous states respecting translational invariance, while on the other hand, there exist real materials, whose ground states are characterised by spontaneously broken translational symmetry and in particular, there are spatially modulated states featured by charge and spin density waves, such as pseudogap regime of cuprate superconductors that breaks translational symmetry and some or all the symmetries of the underlying lattice \cite{vojta}, which calls for similar studies for such kinds of systems.

Studies of holographic models, simulating the effect of breaking of translational symmetry in homogeneous models appear in a number of works. Gravitational solution with homogeneity are often inflicted with instabilities leading to spatially inhomogeneous solutions as demonstrated in various models in bottom-up \cite{ooguri,ooguri1,donos, donos1, donos2, donos3, donos4, cremonini1, cremonini2, cremonini3, cremonini4, withers, rozali, withers1, Withers:2013kva,amoretti, donos5, goute} as well as in top-down approach \cite{bergman, Jokela:2011eb, jokela, jokela1}. In particular, \cite{cremonini3} considered a bottom-up model with two U(1) gauge fields that leads to a pair density wave with coexistence of a superconducting phase and a charge density wave. Similar inhomogeneous solutions are also obtained for Einstein-Maxwell-dilaton model \cite{withers},  axionic system \cite{rozali}, higher derivative gravity model with complex scalar and gauge field\cite{amoretti}  and a checkerboard solution, which breaks translation symmetry spontaneously in two directions \cite{withers1}. In the top-down approach, a D3-D7 model has been found to develop instabilities \cite{jokela} leading to a charge density and a spin density waves \cite{jokela1}. Instabilities of a  similar D2-D8 model have been analysed in \cite{bergman,Jokela:2011eb} and found to lead to a combination of spin and charge density waves.

The Fermi surfaces of such spatially modulated solutions obtained from a holographic perspective, have been analysed and several works have appeared on this score \cite{liu,ling,Cremonini:2018xgj,Balm:2019dxk,Iliasov:2019pav,Cremonini:2019fzz,Andrade:2017ghg}. Periodic lattices were introduced by considering perturbatively small periodic modulation of the chemical potential in \cite{liu}, which was further extended by incorporating the back reaction on the gravity \cite{ling}. These lead to anisotropic Fermi surface as well as a band gap at the boundary of the Brillouin zone, though the lattice periodicity is introduced manually and is irrelevant in the infrared. Fermi surfaces have been studied for striped solution obtained in a bottom-up approach with co-existing charge density wave and superconducting phases \cite{Cremonini:2018xgj}. In addition, they introduced a lattice by periodic modulation of the chemical potential and studied the Fermi surface. They find when the Fermi surface is large enough and cross the Brillouin zone boundary, it develops a gap, which increases as the strength of the lattice increases.

The aim of the present work is to study the Fermi surface of the spatially modulated solution obtained in a D2-D8 model. As mentioned above, instability in a D2-D8 model \cite{bergman,Jokela:2011eb} leads to a combination of a charge density wave (CDW) and a spin density wave (SDW), which was obtained numerically in \cite{Rai:2019qxf}. As has been shown there that this solution is thermodynamically stable with respect to the homogeneous solutions and other instabilities within a domain of the parameter space. 
%

%One can consider the Dirac equation emanating from the top down fermions and use the holographic techniques to study the spectral function of the fermionic operator that couples to the fermions. However, in the action of the D8 brane, the fermions transform in the adjoint representations of the gauge group. Since this solution corresponds to  a single probe D8 brane, the world volume gauge group is U(1) and  the usual coupling of fermions in the fundamental to the gauge field is absent. Nevertheless, we would like to study the behaviours of the Fermi surfaces in the case of this spatially modulated solution which consists of a combination of a charge density wave and a spin density wave.
%

To begin with, we have manually introduced a generic probe fermion in this background coupled to the gauge field in the fundamental. This spatially modulated solution exhibits spontaneous breaking of the translational symmetry. In addition, in order to consider the effect of explicit breaking of the translational symmetry, we have also manually introduced an ionic lattice simulated by choosing a periodically varying chemical potential. 

In the case of generic fermions, we find that the dual theory admits Fermi surface once the charge is sufficiently large and as the fermionic charge increases, the Fermi surface crosses the boundary of the Brillouin zone. At the point where two Fermi surfaces intersect, a gap develops due to Umklapp scattering leading to Fermi pockets (inner part of the Fermi surface). In the presence of an ionic lattice, the gap widens and study of the spectral density function at the inner part indicates that a sufficiently large value of the strength of the ionic lattice will lead to disappearance of Fermi pockets (inner part of the Fermi surface).

Subsequently we study the worldvolume fermions, which couple to the fermionic operators living in the dual $(2+1)$-dimensional Yang-Mills theory. A similar study of top-down fermions appeared in \cite{Ammon:2010pg} in a linearised approximation. We have studied the spectral function associated with the fermionic operators dual to the worldvolume fermions.  However, it does not show a Fermi surface.

We have organised the paper in the following manner. Section 2 consists of a brief review of the charge and spin density wave solution obtained from the D2-D8 model in \cite{Rai:2019qxf}. The third section comprises of a discussion of the Dirac equation for the generic fermions and the numerical techniques employed. In the fourth section we present the results for the generic fermions. We discuss the worldvolume fermions in the fifth section and the sixth section consists of the conclusion.

%%%

%%%%%%%%%%%
			\section{D2-D8 model }
			 The model consists of a configuration of a probe D8 brane along with N D2 branes in the background, which due to instability \cite{Jokela:2011eb,jokela} leads to a spatially modulated solution. In \cite{Rai:2019qxf} a numerical solution ensuing from this instability was obtained, which is characterised by a charge density and a spin density wave and in this section we will review it. If we consider $N$ to be large and the `t Hooft coupling also to be large, $g_sN >> 1$, this gravity theory is holographically dual to a super Yang-Mills (SYM) theory in $(2+1)$ dimensions. In addition, the dual theory also involves charged fermions that follows from the low energy degrees of freedom of bi-fundamental strings. In addition, we will also introduce effect of an underlying lattice by choosing a chemical potential, which is periodic in the $x$ direction, in which the translational symmetry is explicitly broken. 

 The ten-dimensional metric and other fields representing the solution associated with the N D2 branes are given by \cite{Jokela:2011eb,jokela} as follows:
  \begin{equation}\begin{split} \label{metric}
 &\quad  ds^2 = L_0^2 [ r^{5/2} (-f(r) dt^2 + dx^2 + dy^2) + r^{-5/2} (\frac{dr^2}{f(r)} + r^2 dS_6^2)], \\
  dS_6^2 &= d\psi^2 + \sin^2\psi (d\theta_1^2 + \sin^2\theta_1 d\phi_1^2 ) + \cos^2\psi (d\xi^2 + \sin^2\xi d\theta_2^2 + \sin^2\xi \sin^2\theta_2 d\phi_2^2).\\
 f(r) &= 1 - (\frac{r_T}{r})^5, \quad \text{dilaton:} \quad e^\phi = g_s (\frac{r}{L_0})^{-5/4}, \\
 &\text{five-form potential}\quad C^{(5)} = c(\psi) L_0^5~ d\Sigma_2 \wedge d\Sigma_3
 \end{split} \end{equation}
 where $d\Sigma_2$ and $d\Sigma_3$ are the volume forms on the $S^2$ and $S^3$ respectively, $\psi$ takes values from 0 to $\pi/2$, $\xi$, $\theta_1$ and $\theta_2$ take values from 0 to $\pi$ and  $\phi_1$ , $\phi_2$ take values from 0 to 2$\pi$. $c(\psi)$ is given by,
 \begin{equation}
 c(\psi) = \frac{5}{8} \Big(\sin \psi - \frac{1}{6} \sin (3\psi) - \frac{1}{10} \sin{5\psi} \Big),
 \end{equation}
 $L_0$ is given by $L_0^5 = 6 \pi^2 g_s N l_s^5$ and in the following we set $L_0=1$.
  
In this background, there is a probe D8 brane along the directions $t$, $x$, $y$, $r$, $\theta_1$, $\phi_1$, $\xi$, $\theta_2$, $\phi_2$. Its position along $\psi$ direction varies over $r$ and $x$ and the function $\psi(r,x)$ represents the embedding. A magnetic field on $S^2$ given by 
\be\label{magneticfield}
2 \pi \alpha^\prime F_{\theta_1\phi_1} = L_0^2 b \sin\theta_1
\ee 
makes this configuration stable.  There is a gauge field $a_\mu$, with $\mu = (t,x,y,r)$, which depends on the radial direction and $x$ and with a choice of a radial gauge one sets $a_r=0$. One can consider a constant magnetic field $h$ along the $xy$ direction, $2 \pi \alpha^\prime  F_{xy} = h$ which we choose to be zero for present discussion.

The action is given in the following expression,
\begin{equation}\begin{split}
S &= S_1 + S_2,\\
 &= - T_8 \int d^9x ~ e^{-\phi} \sqrt{-det (g_{\mu\nu} + 2 \pi \alpha^\prime F_{\mu\nu})} - \frac{T_8}{2} (2 \pi \alpha^\prime)^2 \int C^{(5)} \wedge F \wedge F,
 \end{split}\end{equation}
 where the first term represents DBI action, while the second one is a Chern-Simons term.

For convenience, we trade $u$ for $r$ where $u=\frac{r_T}{r}$. We also scale the worldvolume coordinates $x_\mu = \hat{x}_\mu r_T^{-3/2}$, the gauge field $a_\mu = r_T \hat{a}_\mu$   and the parameter $b$, $ \hat{b}= b \sqrt{r_T}$.
In terms of $u$ and the rescaled coordinates, the action reduces to
\begin{equation}\begin{split}
S  & = - N r_T^2 \int du~ d\hat{x}~ \frac{1}{u^2}~\sqrt{D (D_1 + D_2 + D_3)}  - N r_T^2 \int du ~ d{\hat x} ~ c(\psi) ({\hat a}_{0u} {\hat a}_{{\hat y}{\hat x}} - {\hat a}_{0{\hat x}} {\hat a}_{{\hat y}u}),\\
  & = - N r_T^2 \int du d{\hat x}~ \frac{1}{u^2} [ \sqrt{D (D_1 + D_2 + D_3)} + u^2 c(\psi) ({\hat a}_{0u} {\hat a}_{{\hat y}{\hat x}} - {\hat a}_{0{\hat x}} {\hat a}_{{\hat y}u}) ] ,
\end{split}\end{equation}
where in the first and second lines the two terms refer to the DBI action and the Chern-Simons term respectively and  $N= 8\pi^3 T_8 V_{1,1}$, $V_{1,1}$ is the volume of spacetime in $t$ and $y$ direction.  $D$, $D_1$,$D_2$ and $D_3$ are defined as follows:
\begin{equation}\begin{split}
D&= \cos ^6{\psi} \left(\sin ^4{\psi}+\frac{{\hat b}^2}{u}\right),\\
D_1 & = \frac{1}{u^5}[ 1 + u^2 f \psi_u^2 -  u^4 {\hat a}_{0u}^2  + u^4 f {\hat a}_{{\hat y}u}^2  + f u^4 {\hat a}_{{\hat x}u}^2] \\
D_2 & = \frac{\psi_{\hat x}^2}{u^2} - \frac{{\hat a}_{0{\hat x}}^2}{f} + {\hat a}_{{\hat y}{\hat x}}^2\\
D_3 &= - u^2 {\hat a}_{0u}^2 \psi_{\hat x}^2     - u^2  {\hat a}_{0{\hat x}}^2 \psi_u^2  + u^2 f {\hat a}_{{\hat y}u}^2 \psi_{\hat x}^2   -  u^4 {\hat a}_{0{\hat x}}^2  {\hat a}_{{\hat y}u}^2   +  u^2 f {\hat a}_{{\hat y}{\hat x}}^2 \psi_u^2   -   u^4 {\hat a}_{0u}^2 {\hat a}_{{\hat y}{\hat x}}^2 \\
& + 2 u^2 {\hat a}_{0u} {\hat a}_{0{\hat x}} \psi_u \psi_{\hat x} + 2 u^4 {\hat a}_{0u} {\hat a}_{0{\hat x}} {\hat a}_{{\hat y}u} {\hat a}_{{\hat y}{\hat x}} - 2 u^2 f {\hat a}_{{\hat y}u} {\hat a}_{{\hat y}{\hat x}} \psi_u \psi_{\hat x},  
\end{split}\end{equation}
where $\psi_u = \frac{\partial \psi}{\partial u}$, ${\hat a}_{0{\hat x}} = \frac{\partial {\hat a_0}}{\partial {\hat x}}$, etc.
Since ${\hat a}_{\hat x}$ decouples from the rest of the system we have dropped it.

The equations of motion ensuing from the above action are nonlinear. Nevertheless, these partial differential equations in ${\hat x}$ and $u$ can be solved using numerical techniques. These equations have the following symmetries: ${\hat x} \rightarrow -{\hat x}$, $\hat{x}\rightarrow \frac{L}{2} - \hat{x}$ and a reflection symmetry for $h=0$.

With a chemical potential $\mu(\hat{x})$, the boundary conditions at the ultraviolet limit, $u=0$ are,
\begin{equation}\begin{split}\label{boundaryuv}
\psi({\hat x},0) &= \psi_\infty ,\quad\quad
\partial_u \psi({\hat x},0) = m_\psi ,\\
{\hat a}_0({\hat x},0) &= \mu({\hat x}), \quad\quad\quad
{\hat a}_y(x,0) = 0,
\end{split}
\end{equation}
 $m_\psi$ represents mass of the fermion, which we have chosen to be non-zero to avoid instabilities arising from tachyons. $\psi_\infty$ is the asymptotic value of the field $\psi$ at $u\rightarrow 0$, which we have chosen to be constant. In order to simulate the effect of a periodic lattice in the ${\hat x}$ direction, following  \cite{Cremonini:2018xgj}, we have introduced a periodic variation of the chemical potential
 \be
 \mu({\hat x}) = \mu (1 + a_i \cos{p{\hat x}}),
 \ee
where $a_i$ represents the relative strength of the one-dimensional lattice and p represents the wavevector associated with the lattice. In the present case, we also have an instrinsic wave vector $K = \frac{2\pi}{L}$. In general, they may be different, but once their values are sufficiently close there will be a commensurate lock-in, which may lead to greater stability \cite{Andrade:2017ghg}. In the present discussion we have chosen them to be equal.

A periodic boundary condition is imposed along ${\hat x}$ direction, so that $\psi$, $a_0$ and $a_y$ are periodic along ${\hat x}$ with periodicity $L$.

At the asymptotic boundary the zeroeth component of the gauge field can be expanded as, 
\begin{equation}
a_0(\hat{x},u) = \mu({\hat x}) + d({\hat x}) u^2 + ..., 
\end{equation}
where $d({\hat x})$ represents the charge density function in the boundary field theory \cite{Jokela:2011eb,jokela}. 
The average of the charge density over the period is given by
\begin{equation}
<d> = \frac{1}{L} \int\limits_0^L d({\hat x}) ~d{\hat x}.
\end{equation}
while the amplitude is Max$(d({\hat x}) - <d>)$. 

Similarly, $\psi$ has the following expansion
\begin{equation}
\psi(\hat{x},u) = \psi_\infty + m_\psi u - c_\psi({\hat x}) u^3 + ....\label{asymbehave}
\end{equation}
with $c_\psi({\hat x})$ representing the fermion bilinear in the dual field theory \cite{Jokela:2011eb,jokela}. $<c_\psi>$ and 
Max($c_\psi({\hat x})-<c_\psi>$) represents the average and the amplitude of the spin density wave in the boundary field theory.

As one may observe, the term  $\frac{(\partial_{\hat x} {\hat a}_0)^2}{f}$ term in $D_2$ diverges at the horizon, since $f$ vanishes there. so we set
\begin{equation}\label{boundaryir}
{\hat a}_0({\hat x},1)=0.
\end{equation}
to have  ${\hat a}$ a well-defined one-form at the horizon.

In order to numerically solve the partial differential equations following from the action one can use pseudospectral method\cite{boyd}. One consider expansion of $\psi$, $a_0$ and $a_y$ along $u$ and ${\hat x}$ direction in terms of suitable functions. Along $u$ and ${\hat x}$ one chooses Chebyshev polynomial and Fourier series respectively.
\begin{equation}
\begin{split}
\psi = \sum\limits_{m=0}^{N_1-1} \sum\limits_{n=0}^{N_2-1} \psi[m,n] T_m(2u-1) \cos \frac{2\pi n \hat{x}}{L} , \\
{\hat a}_0 = \sum\limits_{m=0}^{N_1-1} \sum\limits_{n=0}^{N_2-1} a_0[m,n] T_m(2u-1) \cos \frac{2\pi n \hat{x}}{L} , \\
{\hat a}_y = \sum\limits_{m=0}^{N_1-1} \sum\limits_{n=0}^{N_2-1} a_y[m,n] T_m(2u-1) \sin \frac{2\pi n \hat{x}}{L} . \label{expansion}
\end{split}\end{equation}

The collocation points along $\hat{x}$ direction are distributed uniformly  and the points from $u=0$ to $u=1$ form the Gauss-Lobatto grid. The ultraviolet boundary conditions (\ref{boundaryuv}) reduces  the number of the coefficients. Substitution of these expansions in the equations of motion and evaluating them at collocation points leads to a set of algebraic equations, which can be solved using using Newton-Rhapson method.  The choice in \cite{Rai:2019qxf} is $N_2=9$ and $N_1=11$. 

The values for the various parameters are determined as follows.  As shown in \cite{Jokela:2011eb,jokela}, the asymptotic value of $\psi$ at the boundary $u=0$ should be $\psi_\infty =0$. If at a finite value of $u$, $\psi(u)$ vanishes with finite $\psi^\prime$, self-intersection of D8 brane leads to a conical singularity leading to tachyonic mode, which can be avoided by choosing non-zero $m_\psi$  \cite{Jokela:2011eb}. In \cite{Rai:2019qxf}, $m_\psi$ and b are chosen as $m_\psi=0.5$ $b=1$. 

As explained in \cite{Rai:2019qxf}, with this choice of parameters, one can numerically solve the equations and evaluate the free energy for different values of the chemical potential and periodicity. For a given value of the chemical potential, the periodicity corresponds to the minimum of the free energy. Examining the charge density and the spin density these solutions are found to be characterised by charge density wave and spin density wave. There is a domain in the parameter space over which these spatially modulated charge density wave solutions are thermodynamically stable in comparison to a homogeneous solution. 

In what follows, we have set the magnetic field $h$ to be zero and  the chemical potential $\mu$ deep inside the region of stability.  The periodicity $L$ follows from the minimum of the free energy. With this solution as the background, we will introduce fermions in the gravity theory and consider the Dirac equation in the following section. Solving the Dirac equation numerically we will obtain the spectral density function and study its behaviour.

\section{The Dirac Equation}

Our objective is to study the spectral functions associated with the fermionic operators in the dual theory. In this section we will introduce generic fermions in the gravity theory and study the Dirac equation. The Dirac equation with the generic fermions is given by
\be
[\Gamma^\mu (\nabla_\mu - i q a_\mu) - m ]\chi = 0,
\ee
where the gamma matrices are $\Gamma^\mu = e_a^{~\mu}\Gamma^a$, the covariant derivatives are given by $\nabla_\mu\chi = [\partial_\mu  + \frac{1}{4} (\omega_\mu)_{ab}\Gamma^{ab}]\chi$, q is the charge of the fermions, $a_\mu$ is the background gauge field and m is the mass of the Dirac fermions. We will drop the hats in what follows to avoid the cluttering.

The expressions for the vielbeins $e_a^{~\mu}$ follows from the background metric and for our purpose are given by
\be\begin{split}
e_{\bar t} = \frac{r_T^{1/4} u^{5/4}}{\sqrt{f}} \partial_t,
\quad
e_{\bar x} = \frac{r_T^{1/4} u^{5/4}}{\sqrt{1 + u^3 \psi_x^2}} \partial_x,
\quad
e_{\bar y} = r_T^{1/4} u^{5/4} \partial_y,\\
e_{\bar u} = r_T^{1/4} u^{3/4} \sqrt{f} \sqrt{\frac{ 1 + u^3 \psi_x^2}{1 + u^3 \psi_x^2 + u^2 f \psi_u^2}} ~\Big[ \partial_u - \frac{u^3 \psi_x \psi_u}{1 + u^3 \psi_x^2} \partial_x \Big],
\end{split}\ee
where we have used ${\bar t, \bar x, \bar y, \bar u}$ for the tangent space coordinates, $\psi_u = \frac{\partial \psi}{\partial u}$, $\psi_x = \frac{\partial \psi}{\partial x}$.

We have chosen the following  gamma matrices \cite{liu}, and are written in terms of the Pauli spin matrices as,
\be\begin{split}
\Gamma^{\bar t} &= \left(\begin{array}{cc} i \sigma^1&0\\0&i \sigma^1\end{array}\right),
\quad
\Gamma^{\bar x} = \left(\begin{array}{cc} -\sigma^2&0\\0&\sigma^2\end{array}\right),\\
\Gamma^{\bar y} &= \left(\begin{array}{cc} 0&-i \sigma^2\\ i \sigma^2&0\end{array}\right),
\quad
\Gamma^{\bar u} = \left(\begin{array}{cc}  -\sigma^3&0\\0&-\sigma^3\end{array}\right).
\end{split}
\ee

The spin connection $(\omega_\mu)_{ab}$ can be absorbed by making the following redefinition of the spinors
\be
\chi = \frac{ r_T^{3/8} u^{5/8}}{ (1 + u^3 \psi_x^2)^{1/4}} \left(\begin{array}{c} \Psi_1\\ \Psi_2\end{array}\right),
\ee
where $\Psi_\alpha, \alpha=1,2$ is a two-component spinor. Since the background is spatially modulated in the $x$ direction with a period fixed by Umklapp wavevector $K=\frac{2 \pi}{L}$, the momentum modes that differ by a lattice vector are not independent. In accordance with  the Bloch Theorem, we consider the following expansion,
\be
\Psi_\alpha = \int \frac{d\omega dk_x dk_y}{2\pi} \sum\limits_{l\in Z} {\mathcal F}_\alpha^{(l)} ( u, \omega,k_x,k_y) e^{-i\omega t + i (k_x + l K)x + i k_y y },\quad \alpha=1,2.
\ee
Here l refers to the different momentum level and $k_x$ is restricted to the first Brillouin zone.  We write
\be
{\mathcal F}_\alpha (u, x, \omega, k_y) = \sum\limits_{l\in Z} {\mathcal F}_\alpha^{(l)} ( u, \omega,k_x,k_y) e^{ i l K x },
\ee 
where ${\mathcal F}_\alpha $ satisfy
\be
{\mathcal F}_\alpha (u, x, \omega, k_y) = {\mathcal F}_\alpha (u, x + \frac{2\pi}{K}, \omega, k_y) 
\ee
Since $\Psi_\alpha$ is a two component spinor, we further split ${\mathcal F}_\alpha (u, x)$ as
\be
{\mathcal F}_\alpha (u, x) = \left(\begin{array}{c} {\mathcal A}_\alpha (u,x) \\ {\mathcal B}_\alpha (u,x)\end{array}\right),\quad \alpha = 1,2.
\ee
With these splitting, the Dirac equation can be written as,
\be\begin{split}
\partial_u & \left(\begin{array}{c} {\mathcal A}_1 \\ {\mathcal B}_1 \end{array}\right)
- \frac{\sqrt{u}}{f} \sqrt{\frac{1 + u^3 \psi_x^2 + u^2 f \psi_u^2}{1+u^3\psi_x^2}} (\omega + q a_0)  \left(\begin{array}{c} {\mathcal B}_1 \\ - {\mathcal A}_1 \end{array}\right) 
- \frac{u^3 \psi_x \psi_u}{1+u^3 \psi_x^2}(\partial_x + i k_x ) \left(\begin{array}{c} {\mathcal A}_1 \\  {\mathcal B}_1 \end{array}\right) \\
- i& \sqrt{\frac{u}{f}} \frac{\sqrt{1 + u^3 \psi_x^2 + u^2 f \psi_u^2}}{1+u^3\psi_x^2} (\partial_x + i k_x ) \left(\begin{array}{c} {\mathcal B}_1 \\  {\mathcal A}_1 \end{array}\right) 
+ \frac{\tilde{m}}{u^{3/4}\sqrt{f}}\sqrt{\frac{1 + u^3 \psi_x^2 + u^2 f \psi_u^2}{1+u^3\psi_x^2}}\left(\begin{array}{c} {\mathcal A}_1 \\ - {\mathcal B}_1 \end{array}\right) \\
+ i& \sqrt{\frac{u}{f}} \sqrt{\frac{1 + u^3 \psi_x^2 + u^2 f \psi_u^2}{1+u^3\psi_x^2}} (k_y - q a_y)  \left(\begin{array}{c} {\mathcal B}_2 \\  {\mathcal A}_2 \end{array}\right) = 0 ,\\
%%%%%%%%%%%%%%%%
\partial_u & \left(\begin{array}{c} {\mathcal A}_2 \\ {\mathcal B}_2 \end{array}\right)
- \frac{\sqrt{u}}{f} \sqrt{\frac{1 + u^3 \psi_x^2 + u^2 f \psi_u^2}{1+u^3\psi_x^2}} (\omega + q a_t)  \left(\begin{array}{c} {\mathcal B}_2 \\ - {\mathcal A}_2 \end{array}\right) 
- \frac{u^3 \psi_x \psi_u}{1+u^3 \psi_x^2}(\partial_x + i k_x ) \left(\begin{array}{c} {\mathcal A}_2 \\  {\mathcal B}_2 \end{array}\right) \\
+ i& \sqrt{\frac{u}{f}} \frac{\sqrt{1 + u^3 \psi_x^2 + u^2 f \psi_u^2}}{1+u^3\psi_x^2} (\partial_x + i k_x ) \left(\begin{array}{c} {\mathcal B}_2 \\ {\mathcal A}_2 \end{array}\right) 
+ \frac{\tilde{m}}{u^{3/4}\sqrt{f}}\sqrt{\frac{1 + u^3 \psi_x^2 + u^2 f \psi_u^2}{1+u^3\psi_x^2}}\left(\begin{array}{c} {\mathcal A}_2 \\ - {\mathcal B}_2 \end{array}\right) \\
- i& \sqrt{\frac{u}{f}} \sqrt{\frac{1 + u^3 \psi_x^2 + u^2 f \psi_u^2}{1+u^3\psi_x^2}} (k_y - q a_y)  \left(\begin{array}{c} {\mathcal B}_1 \\  {\mathcal A}_1 \end{array}\right) = 0 ,
\label{dirac}
\end{split} \ee
where we use $\tilde{m} = \frac{m}{r_T^{1/4}}$.
The momentum mode expansion of the different functions ${\mathcal A}_\alpha$ and  ${\mathcal B}_\alpha$ are given by
\be
{\mathcal A}_\alpha  = \sum\limits_{l\in Z} {\mathcal A}_\alpha^{(l)}  e^{ i l K x },\quad
{\mathcal B}_\alpha = \sum\limits_{l\in Z} {\mathcal B}_\alpha^{(l)}  e^{ i l K x },
\ee 

The boundary condition for solving these first order linear differential equations can be obtained from the near horizon limit $u\rightarrow 1$. After substituting the expressions for the background metric and gauge fields and choosing the near horizon limit one finds that the different momentum modes are satisfying the following conditions in the leading order,
\be
{\mathcal A}_\alpha^{(l)} \sim (1-u)^{\pm \frac{i\omega}{5}} a_{\alpha 0}^{(l)},\quad 
{\mathcal B}_\alpha^{(l)} \sim (1-u)^{\pm \frac{i\omega}{5}} b_{\alpha 0}^{(l)}.\label{bc1}
\ee
We have chosen the minus sign and impose the in-falling boundary conditions \cite{Iqbal:2009fd} as that is the correct choice for holographic computation of retarded Green's function of the dual theory living at the boundary. Furthermore, the equations at the near horizon limit also implies
\be 
b_{\alpha 0}^{(l)}  = - i a_{\alpha 0}^{(l)} .\label{bc2}
\ee

In the asymptotic limit near the boundary $u\rightarrow 0$ the Dirac equations (\ref{dirac}) reduces to up to the leading order,
\be
\partial_u  \left(\begin{array}{c} {\mathcal A}_\alpha \\ {\mathcal B}_\alpha \end{array}\right) + \tilde{m} u^{-3/4} \left(\begin{array}{c} {\mathcal A}_\alpha \\ - {\mathcal B}_\alpha \end{array}\right) = 0,
\ee
which implies the asymptotic behaviour of the fermions are given by
\be
{\mathcal F}_\alpha =  \left(\begin{array}{c} {\mathcal A}_\alpha \\ {\mathcal B}_\alpha \end{array}\right) \sim a_\alpha e^{4 {\tilde m} u^{1/4}}  \left(\begin{array}{c} 1 \\ 0 \end{array}\right) + b_\alpha e^{- 4 {\tilde m} u^{1/4}}  \left(\begin{array}{c} 0 \\ 1 \end{array}\right).
\ee
This asymptotic behaviour in terms of the momentum level function becomes
\be
{\mathcal F}_\alpha^{(l)} =  \left(\begin{array}{c} {\mathcal A}_\alpha^{(l)} \\ {\mathcal B}_\alpha^{(l)} \end{array}\right) \sim a_\alpha^{(l)} e^{4 {\tilde m} u^{1/4}}  \left(\begin{array}{c} 1 \\ 0 \end{array}\right) + b_\alpha^{(l)} e^{- 4 {\tilde m} u^{1/4}}  \left(\begin{array}{c} 0 \\ 1 \end{array}\right).
\ee
The retarded Green's function can be obtained from the relation between $a_\alpha^{(l)}$ and $b_\alpha^{(l)}$
as
 \be
 a_\alpha^{(l)}(\omega, k_x, k_y) = \sum\limits_{\alpha^\prime, l^\prime} G^R_{\alpha , l ; \alpha^\prime , l^\prime } (\omega, k_x, k_y)  b_{\alpha^\prime}^{(l^\prime)}(\omega, k_x, k_y), \label{green}
 \ee
The Green's function is considered in the momentum basis as in the experiments such as ARPES, the photoelectrons are in the definite states of momentum. Following \cite{Cremonini:2018xgj} we assume that the dominant response will be in the diagonal momentum channel, though there will be a mixing with other momentum modes. With this assumption the spectral density function can be written as
\be
A(\omega,k_x,k_y) = \sum\limits_{l\in Z} Im (Tr( G^R_{\alpha , l ; \alpha^\prime , l } (\omega, k_x, k_y))),
\label{spF}
\ee
where $-\frac{K}{2} \leq k_x \leq \frac{K}{2}$ is chosen to be within the first Brillouin zone and l denotes the momentum level or Brillouin zone.

In order to compute the Green's function, we will follow the method explained in \cite{liu}. We denote the boundary conditions to be $(\alpha, l)$ by imposing in-falling boundary condition on the spinor component $\Psi_\alpha^{(l)}$ and setting all other spinor components to be zero. We will write the solution for $a_\alpha^{(l)}$ with $(\beta , k)$ boundary condition as $a (\alpha , l ; \beta, k )$ and in this notation the expression for Green's function (\ref{green}) can be written as 
 \be
 a(\alpha ,l ; \beta , k ) = \sum\limits_{\alpha^\prime, l^\prime} G^R(\alpha , l ; \alpha^\prime , l^\prime )  b(\alpha^\prime, l^\prime ; \beta, k). \label{green1}
 \ee
 Writing $a(\alpha ,l ; \beta , k )$ and $b(\alpha^\prime ,l^\prime ; \beta , k )$ as matrices {\bf a} and {\bf b}, the relation becomes
 \be
 {\bf a} = G^R . {\bf b}, \quad\quad   G^R = {\bf a}.{\bf b}^{-1}
 \ee
 Then the spectral weight follows from the Green's function $G^R$.
 
We would like to study the spectral function associated with the generic fermions for the charge density and spin density wave background, which was derived in \cite{Rai:2019qxf}. The study of the generic fermions will give us more flexibility and in order to keep the analysis simple we will further restrict ourselves, to the case of massless fermions. Since we do not have an analytic solution for that we will employ a numerical procedure to solve the Dirac equations for different in-falling boundary conditions and from that we will get the spectral density function in the next section. We will consider the fermions in the probe limit and will not consider the back reaction on the gravity.

\section{Results}

In this section we will consider the spectral function obtained for various parameters in order to study the Fermi surface. At zero temperature, the Fermi surface appears as a pole of spectral density function. In the present case,  we will assume that the Fermi surface will reveal itself through peaks of the spectral density function \cite{Cosnier-Horeau:2014qya}. 

Following \cite{Cosnier-Horeau:2014qya} we assume the plot of the spectral density function vs. momentum at $\omega\rightarrow 0$ will show a peak at around that value of the momentum ${\bf k_0}$, which corresponds to a Fermi surface. The width of the peak should be of the order of the temperature or less. And finally, at that value of the momentum ${\bf k}= {\bf k}_0$, plotting spectral density vs. frequency $\omega$ will show a peak near $\omega = 0$.

With the above criteria, we study the existence and the features of the Fermi surface in the charge and spin density wave solution that we obtained in the present case of D2-D8 brane. We will begin with the solution in absence of any lattice. Due to spontaneous breakdown of the translational symmetry the solutions are characterised with a natural length scale associated with the periodicity of the charge density wave. Given a chemical potential, the free energy of the solution varies with the periodicity and the latter is determined by the minimum of the free energy. In the present case, we have chosen the value of the chemical potential to be $\mu$ = 1.6445, which is deep inside the region of stability of the charge density wave\cite{Rai:2019qxf}. The value of the period, at that chemical potential for $a_i=0$, it turns out to be $L=0.3725$, implying the Bloch wave vector is given by $K=16.8676$. So the first Brilluoin zone lies between $\pm 8.331$. We have also considered non-zero values of the strength of the lattice and keeping the chemical potential same the periodicity turns out to be $L=0.3155$ for $a_i=0.2$. Since the chemical potential remains fixed throughout the discussion, the variables can be measured in that unit, which will lead to a rescaling.

\begin{figure}[h]
			\centering
			\begin{subfigure}{7.5cm}
				\centering
				\includegraphics[width=7cm]{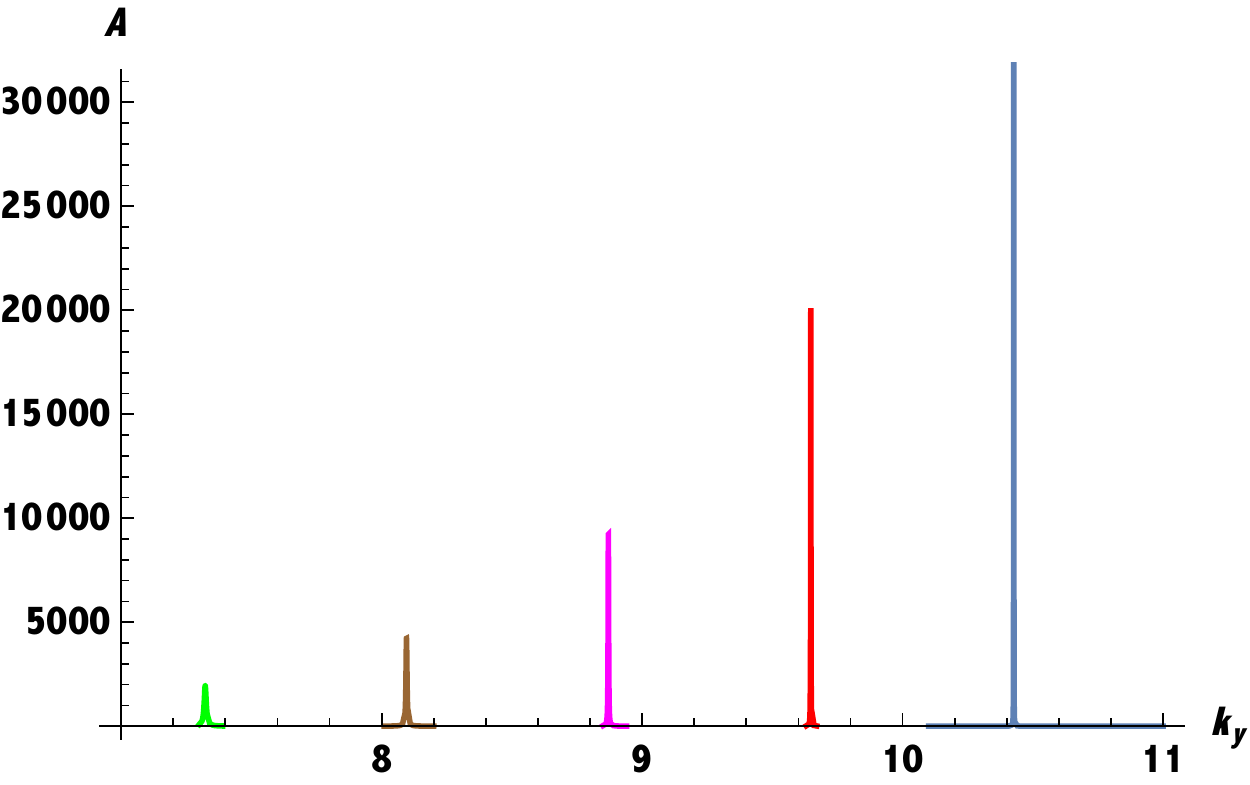}
						\caption{}
				\label{fig:1a}
			\end{subfigure}
			\begin{subfigure}{7.5cm}
				\centering
				\includegraphics[width=7cm]{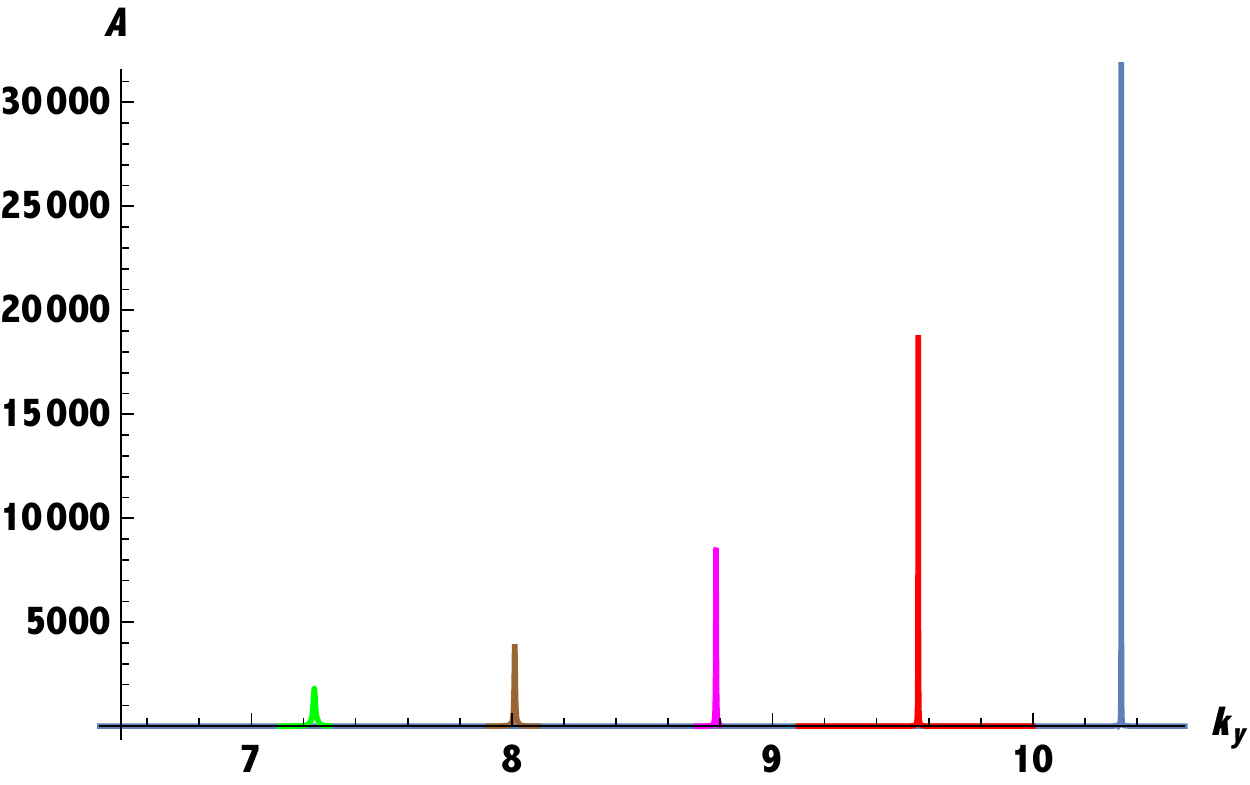}
						\caption{}
				\label{fig:1b}
			\end{subfigure}
			\caption{Plot of spectral function A vs. $k_y$ with $k_x=0$. The peak shifts from right to left with decreasing charges. We have plotted for  $q$=9 (blue), 8.5 (red), 8 (magenta), 7.5 (brown), 7 (green). (a) $a_i=0$ (b) $a_i$=0.2. The height of the plot for $q=9$ is truncated so as to get all the peaks visible in the figure.}
			\label{Avsky}
		\end{figure}

 To begin with we consider how the Fermi surface depends on the charges of the fermions. We choose a specific value of $k_x$ to be $k_x=0$,  a small value of the frequency $\omega$ to be $\omega=.001$ and plotted the spectral function $A$ vs. $k_y$ for five different values of charges. The plots are given in the Figure.\ref{Avsky}(a) in absence of the ionic lattice ($a_i=0$) and in the Figure.\ref{Avsky}(b) in presence of the ionic lattice ($a_i=0.2$). 
 
 As one can see from the figures, with the increase of the fermionic charge the peaks are getting sharper and the heights are increasing. This is a general feature of the holographic fermions which holds here as well. With charge $q$ sufficiently less, the peak will become quite shallow and broad indicating absence of the Fermi surface. As the charge increases, position of the peak in $k_y$ moves to the right, in the positive direction. Since $k_x=0$ it can be associated with the fact that the size of the Fermi circle gets bigger with the increasing charges. 

\begin{figure}[h]
			\centering
			%\begin{subfigure}{8cm}
			%	\centering
				\includegraphics[width=12cm]{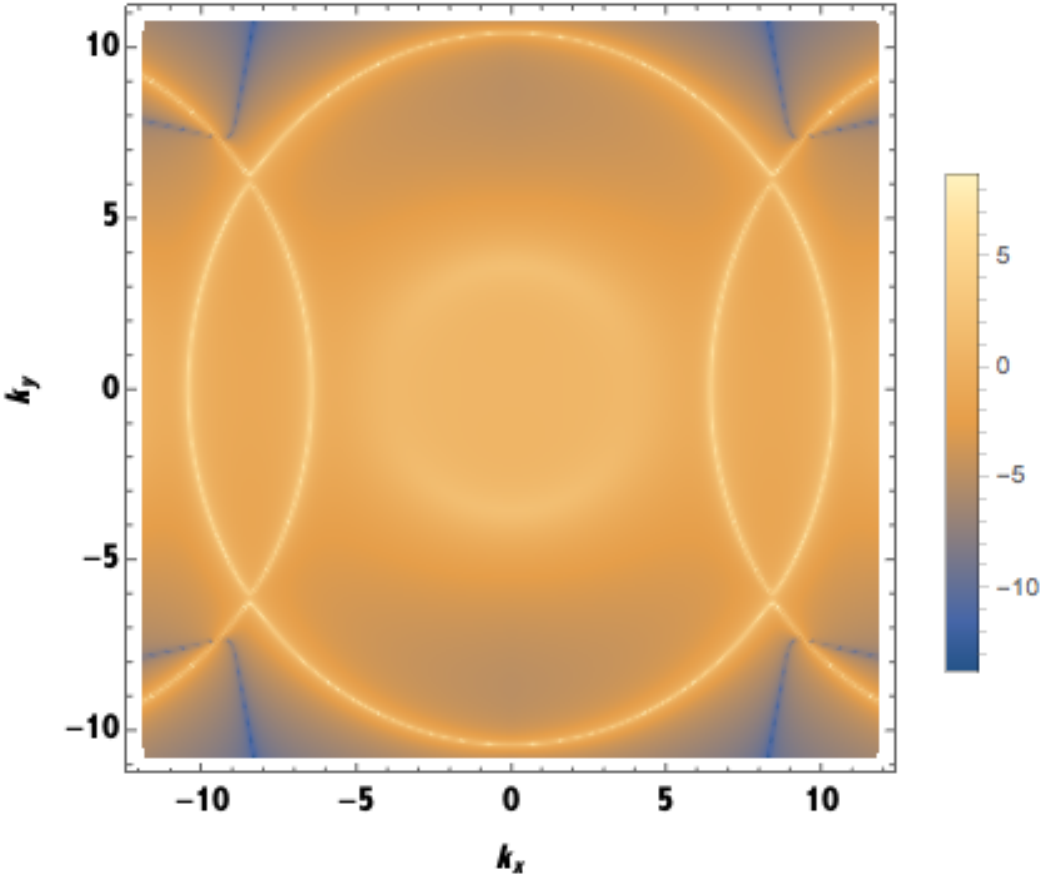}
						%\caption{}
				%\label{fig:1a}
			%\end{subfigure}%
			%\begin{subfigure}{8cm}
			%	\centering
			%	\includegraphics[width=7cm]{3Dfsion}
			%			\caption{}
			%	\label{fig:1b}
			%\end{subfigure}
			\caption{Density Plot of spectral function A over $(k_x, k_y)$-plane in absence of ionic lattice.}
			\label{Avskxky}
		\end{figure}

One important topic of interest of the present work is to examine the existence and behaviour of the Fermi surface. We will begin with the model in absence of the ionic lattice. For the given periodicity we have chosen the fermionic charge to be $q=9$ so that the Fermi surface occurs near the boundary of the Brillouin zone. We have given a density plot in the Figure.\ref{Avskxky}, where we have plotted the spectral function vs. $k_x$ and $k_y$. Since the height of the peaks varies considerably over the region we have given a logarithmic plot of the spectral function. We have numerically computed the values for the first Brillouin zone $-\frac{\pi}{K} \leq k_x \leq \frac{\pi}{K}$ and extended it periodically over $k_x$. 

As one can observe that the Fermi surface consists of intersecting circles with a small eccentricity. At the edge of the Brillouin zone, due to eigenvalue repulsion of degenerate eigenvalues, it opens up a gap in the spectral density. A careful observation would reveal small gaps at the points where the circles intersect as a consequence of the broken translational invariance. The density plot also shows a circle around the origin deep inside the Brillouin zone. With the charge $q=9$, the height and sharpness of the associated peaks do not qualify to be a Fermi surface. However, with increasing charge it will lead to another Fermi surface. Such nested Fermi surface has been shown in \cite{DeWolfe:2012uv}.

\begin{figure}[h]
			\centering
			%\begin{subfigure}{8cm}
			%	\centering
				\includegraphics[width=12cm]{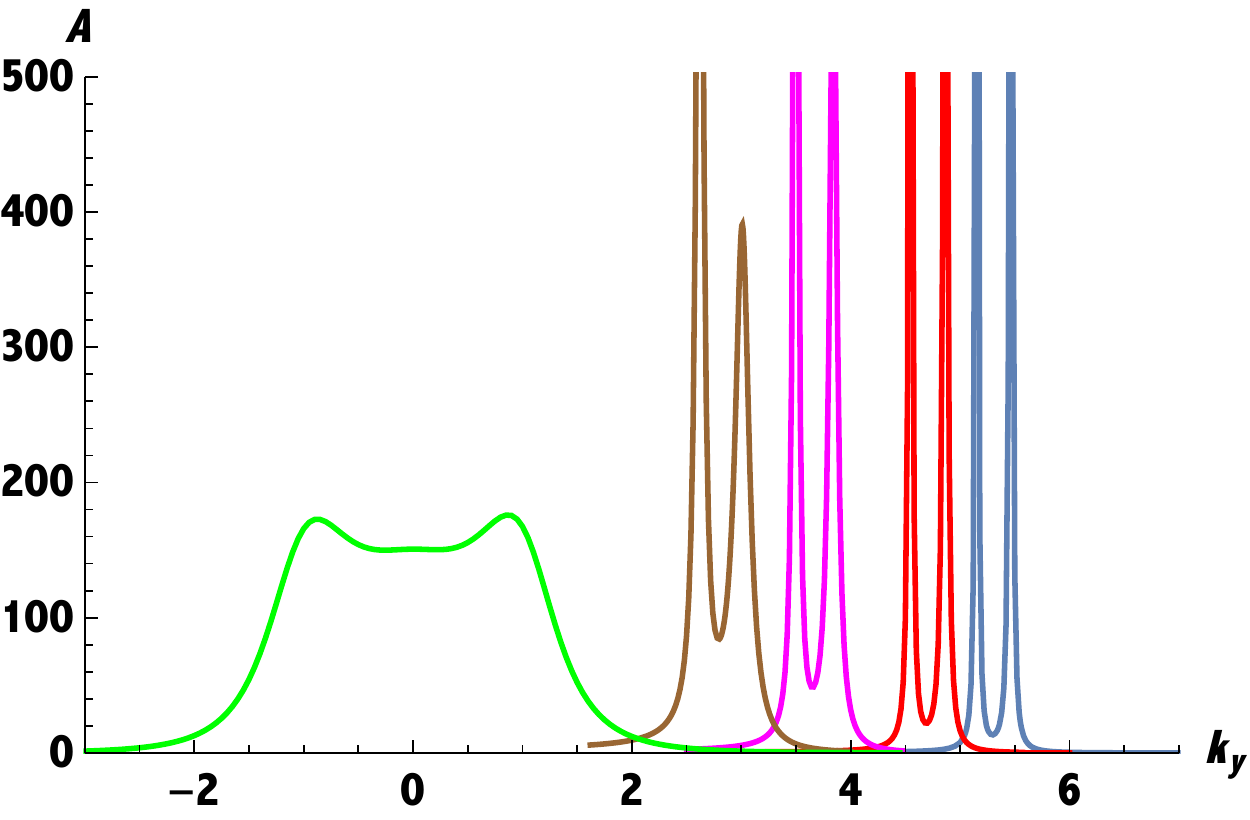}
						%\caption{}
				%\label{fig:1a}
			%\end{subfigure}%
			%\begin{subfigure}{8cm}
			%	\centering
			%	\includegraphics[width=7cm]{3Dfsion}
			%			\caption{}
			%	\label{fig:1b}
			%\end{subfigure}
			\caption{ Plot of spectral function A vs. $k_y$ with $k_x=\frac{\pi}{L}$ at the boundary of the Brillouin zone in absence of ionic lattice with different fermionic charges. $q=$ 8.7 (blue), 8.5 (red), 8.2 (magenta), 8 (brown), 7.7 (Green). The height of some of the plots are truncated so as to get all the peaks visible in the figure.}
			\label{bifurcate}
		\end{figure}

Since the gaps are not very pronounced in the density plot given in the Figure.\ref{Avskxky}, we have plotted the spectral function vs. $k_y$ at the boundary of the Brillouin zone at $k_x=\frac{\pi}{L}$ in the Figure.\ref{bifurcate}. In order to see the variation of the gap we have computed them for several values of charges as shown in the Figure.\ref{bifurcate}. For larger fermionic charges at the boundary, it shows two adjacent peaks in the value of $A$. As the charge decreases, the heights of the two adjacent peaks decreases and the depth of the intermediate region becoming swallow and thus blurring the gap. For $k_x=0$ also the heights are decreasing as the charge decreases, but those are sharp enough to be associated with Fermi surface. This seems to indicate that for sufficiently small charge the Fermi surface will be depleted near the intersection.

\begin{figure}[h]
			\centering
			%\begin{subfigure}{8cm}
			%	\centering
				\includegraphics[width=12cm]{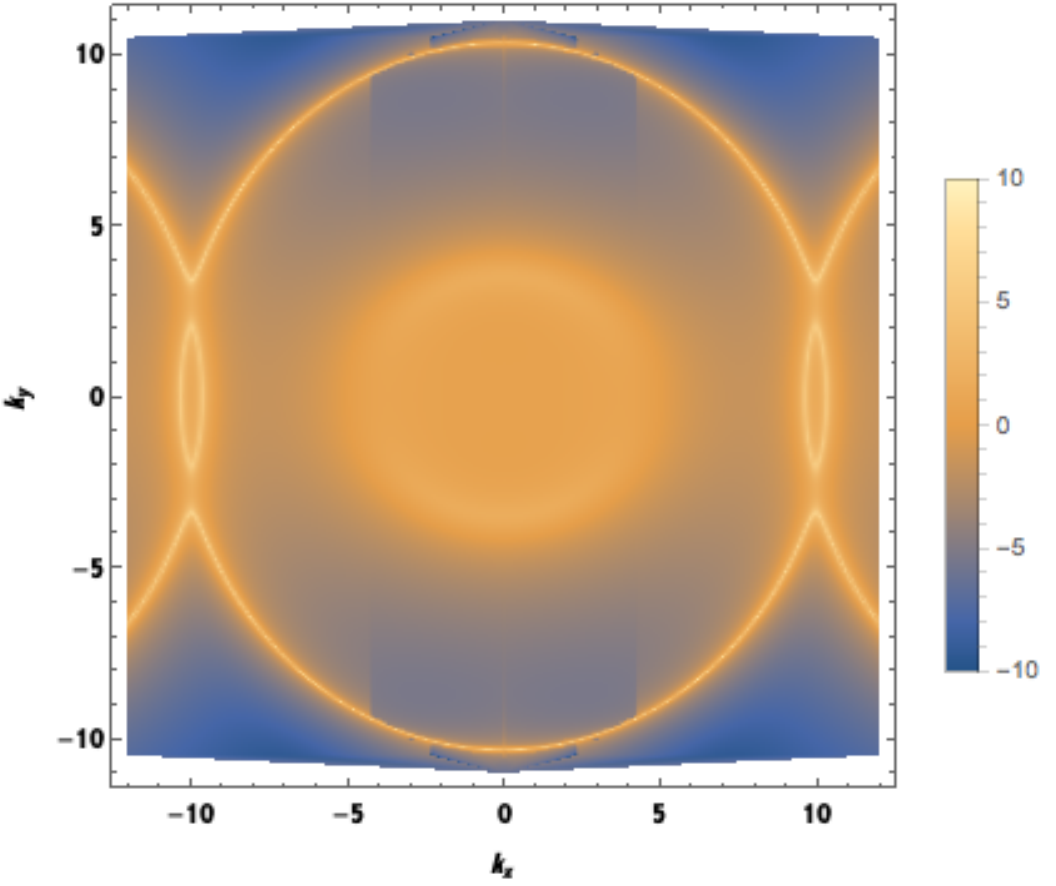}
						%\caption{}
				%\label{fig:1a}
			%\end{subfigure}%
			%\begin{subfigure}{8cm}
			%	\centering
			%	\includegraphics[width=7cm]{3Dfsion}
			%			\caption{}
			%	\label{fig:1b}
			%\end{subfigure}
			\caption{Density Plot of spectral function A over $(k_x, k_y)$-plane in presence of the ionic lattice with $a_i=0.2$.}
			\label{Avskxkyionic}
		\end{figure}

\begin{figure}[h]
			\centering
			%\begin{subfigure}{8cm}
			%	\centering
				\includegraphics[width=12cm]{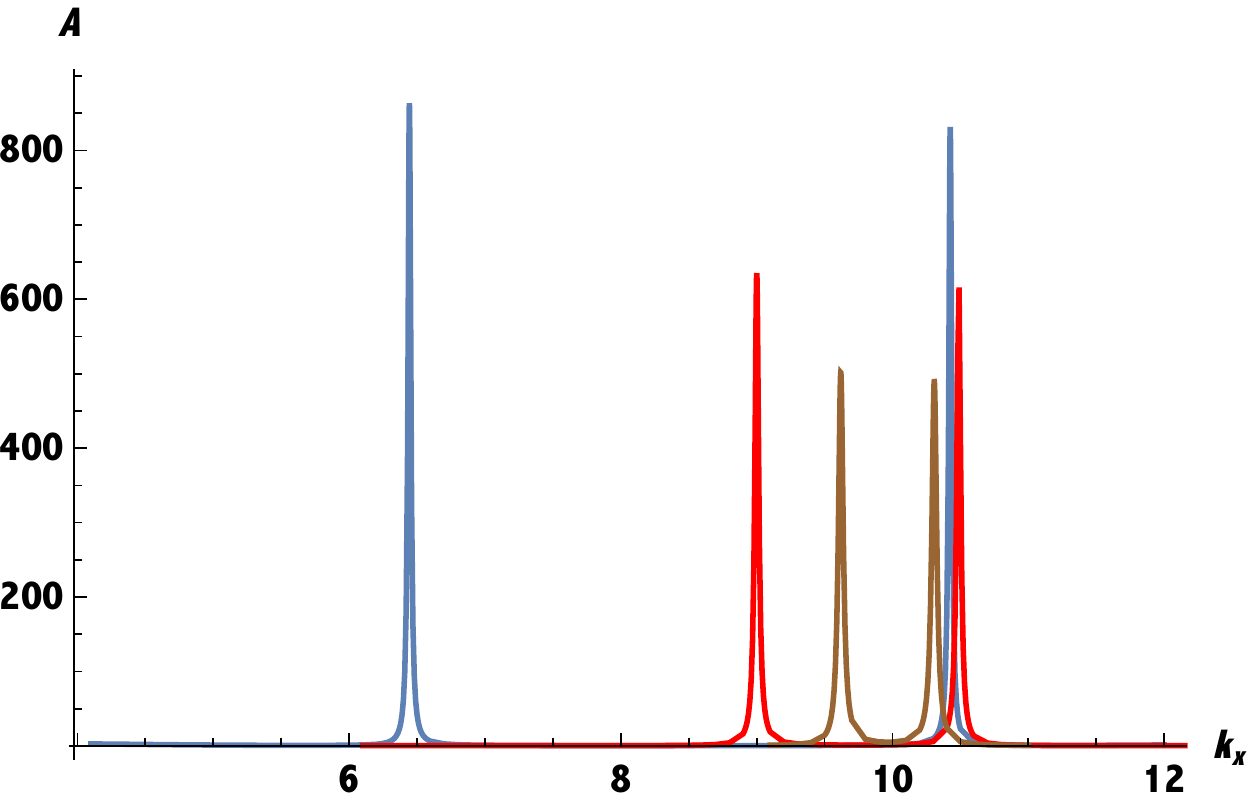}
						%\caption{}
				%\label{fig:1a}
			%\end{subfigure}%
			%\begin{subfigure}{8cm}
			%	\centering
			%	\includegraphics[width=7cm]{3Dfsion}
			%			\caption{}
			%	\label{fig:1b}
			%\end{subfigure}
			\caption{ Plot of spectral function A vs. $k_x$ with $k_y= 0$ with different strengths of ionic lattices $a_i=$ 0 (blue), 0.1 (red), 0.2 (brown).}
			\label{Avskdpl}
		\end{figure}

Next we have turned on the ionic lattice and numerically computed the spectral function, which has been given in a density plot in the Figure.\ref{Avskxkyionic}.  We have chosen $q=9$ once again, and set the strength of the ionic lattice to be $a_i=0.2$. Once again we have obtained intersecting circular Fermi surfaces, which intersects near the boundary of the Brillouin zone. But this time the gaps are wider and It shows small elliptical shaped Fermi pockets near the boundary of the Brillouin zone, which are gapped due to the Umklapp scattering. The gap increases with increasing strength of the ionic lattice, making the Fermi pockets smaller. We have further plotted the spectral function $A$ vs. $k_x$ at $k_y=0$ in the Figure.\ref{Avskdpl} for increasing values of the strength of the ionic lattice from 0 to 0.2. We have chosen $k_y=0$ as it corresponds to substantial peak associated with the Fermi pockets. One can observe, the height of the spectral function is decreasing with the increasing strength of ionic lattice. This indicates, for a sufficiently large value of $a_i$, the strength of the ionic lattice, the Fermi pocket will disappear. 

\begin{figure}[h]
			\centering
			\begin{subfigure}{8cm}
				\centering
				\includegraphics[width=7cm]{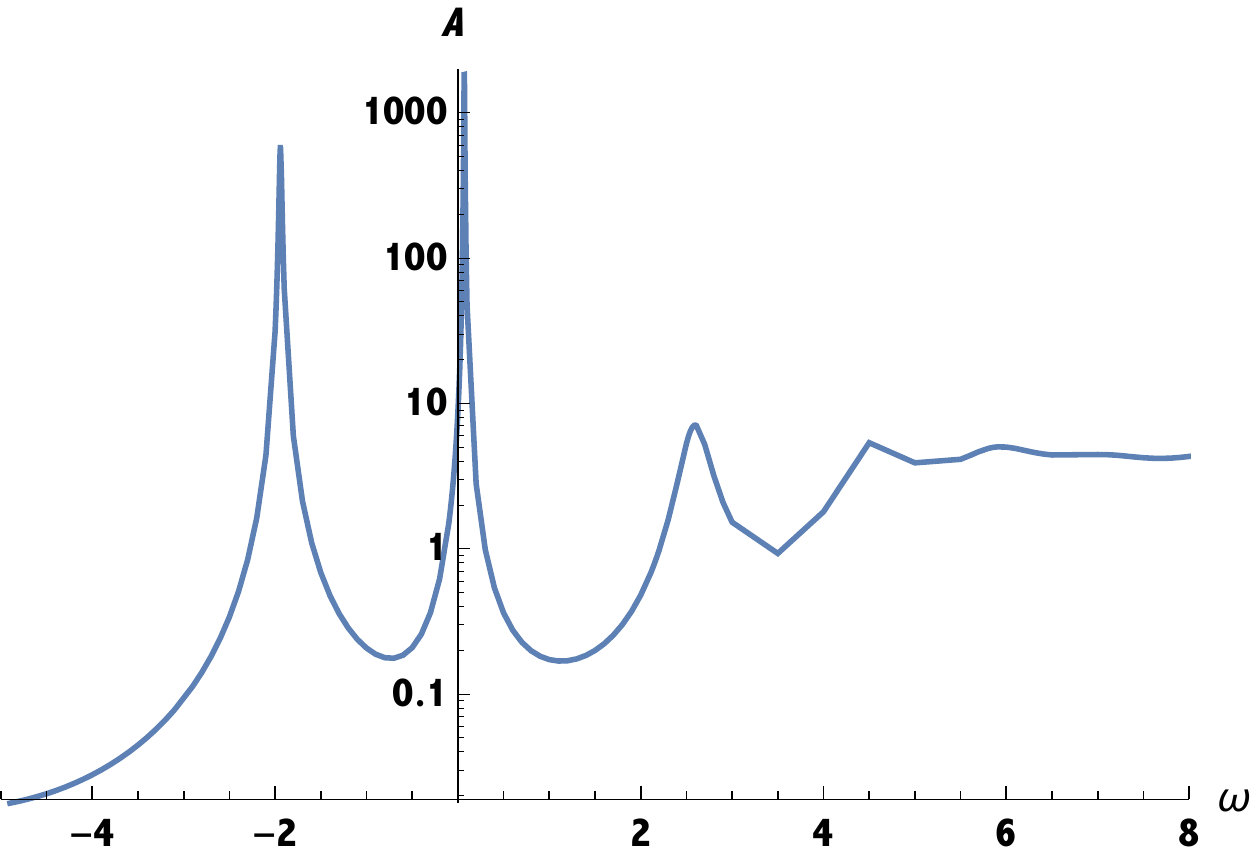}
						\caption{}
				%\label{fig:1a}
			\end{subfigure}%
			\begin{subfigure}{8cm}
				\centering
				\includegraphics[width=7cm]{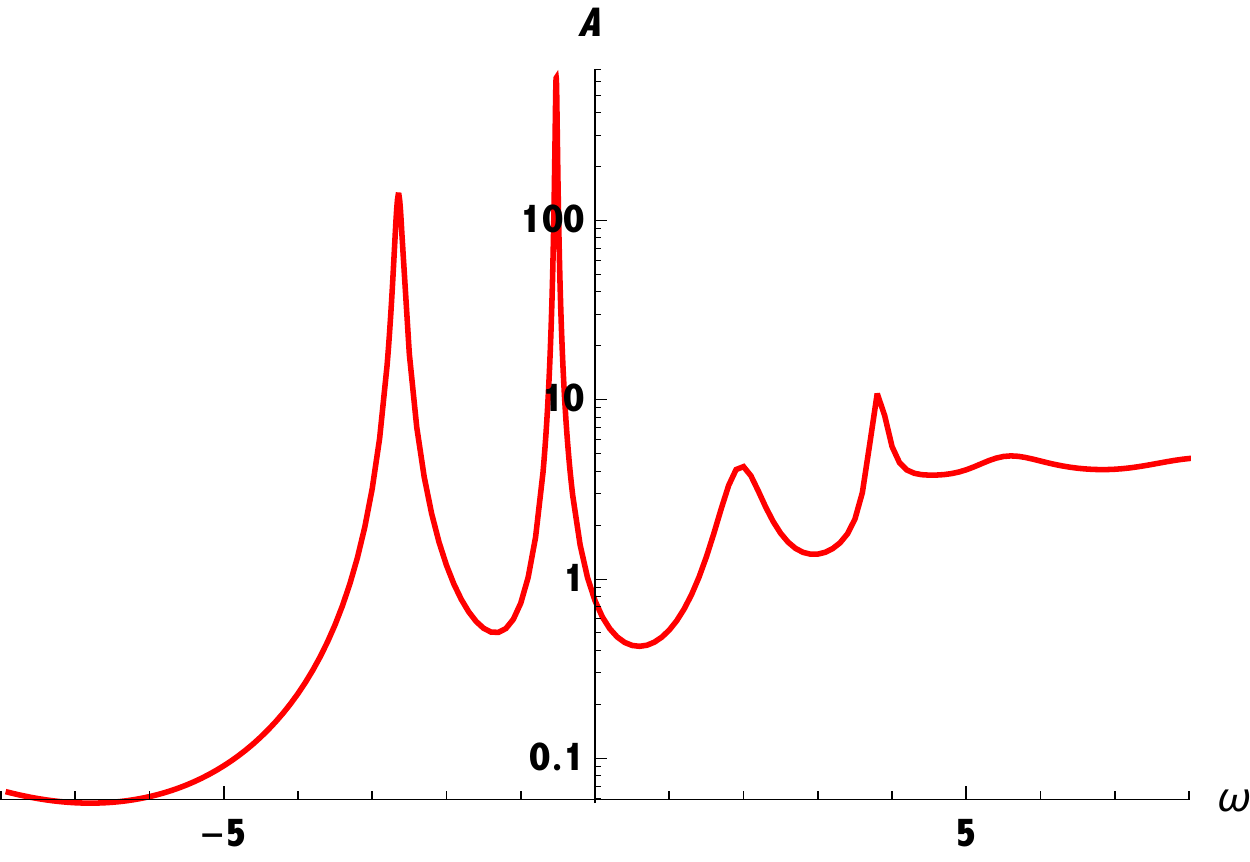}
						\caption{}
			%	\label{fig:1b}
			\end{subfigure}
			\begin{subfigure}{8cm}
				\centering
				\includegraphics[width=7cm]{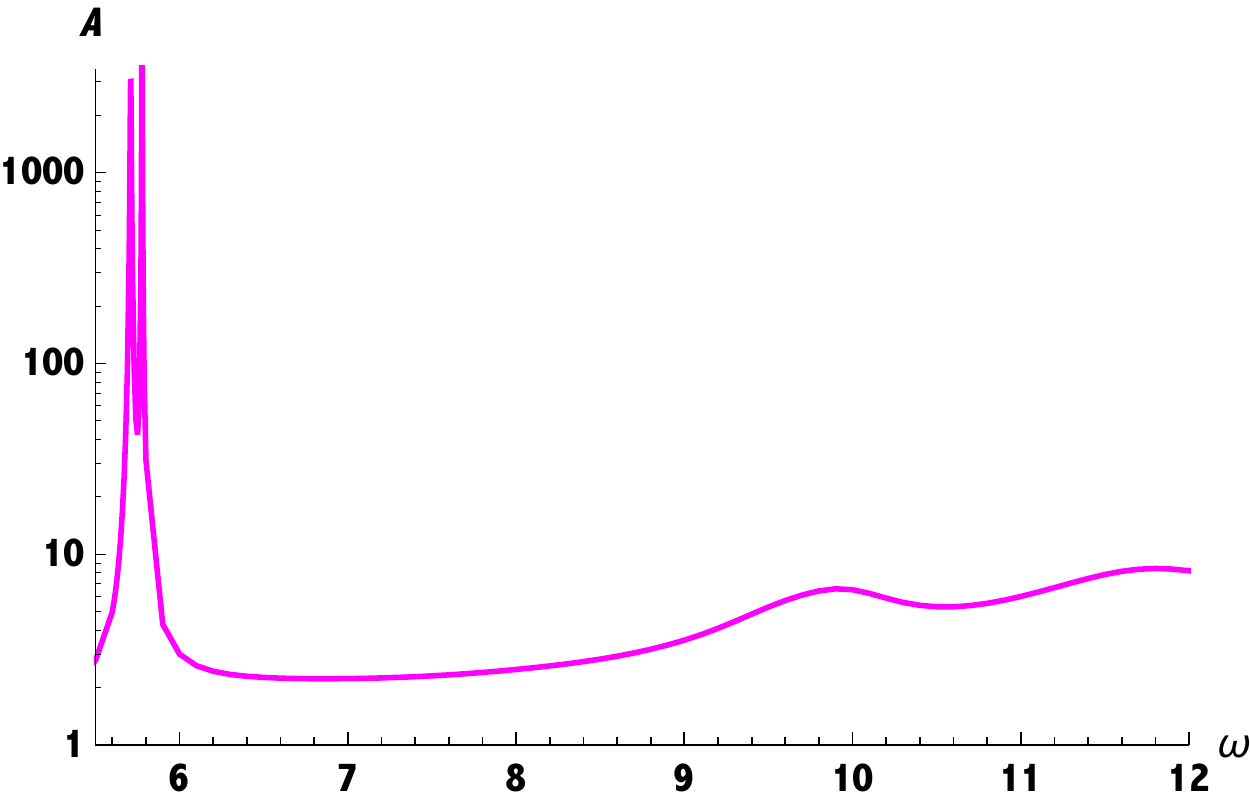}
						\caption{}
				%\label{fig:1a}
			\end{subfigure}%
			\begin{subfigure}{8cm}
				\centering
				\includegraphics[width=7cm]{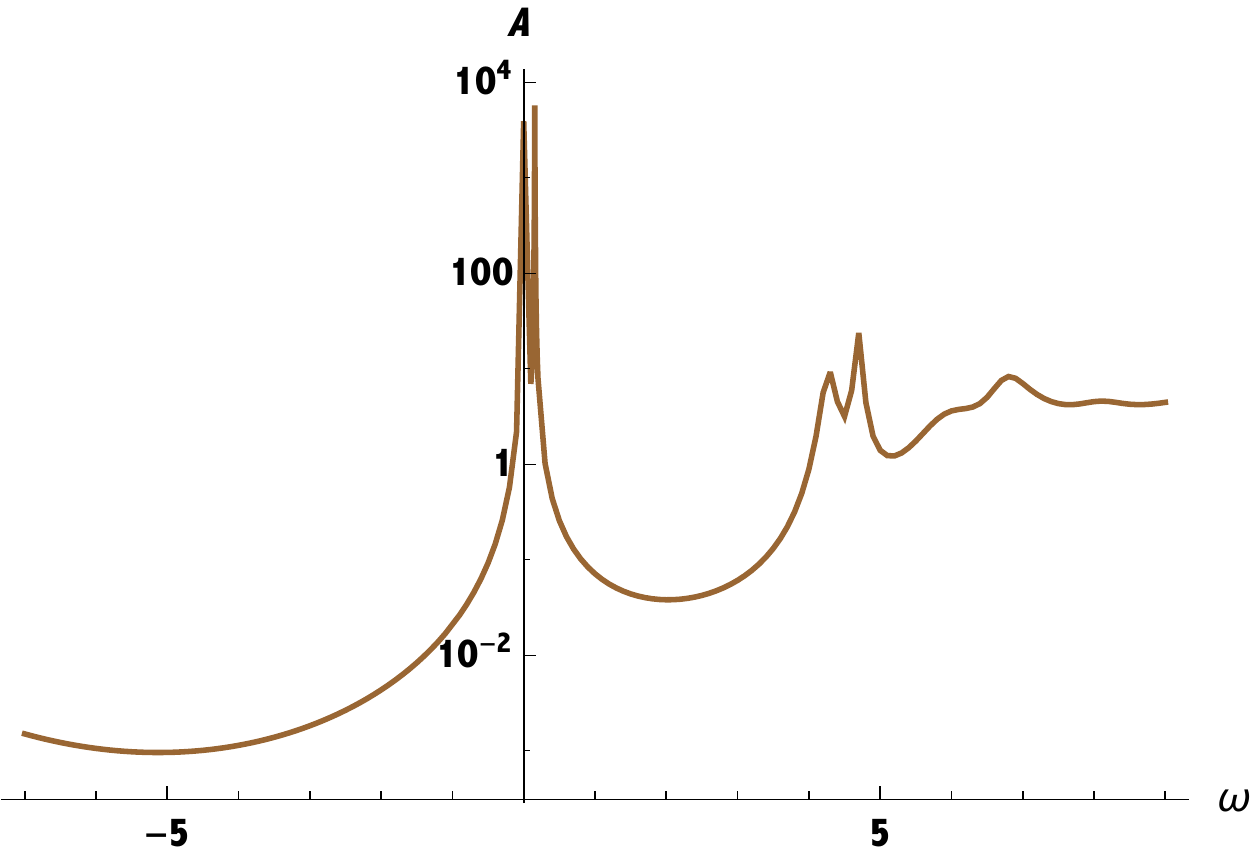}
					\caption{}
			%	\label{fig:1b}
			\end{subfigure}
			\caption{Plot of spectral function A vs. $w$ with $q=9$ in absence of ionic lattice; (a) $k_x=9.8, k_y=3.8$ (blue), (b) $k_x=9.8, k_y=0$ (red), (c) $k_x=0, k_y=0$ (magenta), (d) $k_x=8.43381, k_y=6.287$ (brown)}
			\label{Avsw}
		\end{figure}

In order to study the behaviour of the energy distribution function, we have plotted the spectral function with variation of the frequency, ($A$ vs. $\omega$) in the Figure.\ref{Avsw}. The Figure.\ref{Avsw}(a) and \ref{Avsw}(b) correspond to the plots of spectral function outside and inside the Fermi surface near the boundary. As one can see the peak in $A$ lies at $\omega \ge 0$ and $\omega \le 0$, respectively, while exactly at the Fermi surface position of the peak coincides with $\omega=0$, as expected from the criterion of the Fermi surface. Figure.\ref{Avsw}(c) describes the behaviour of the spectral function deep inside the Brillouin zone and far from the Fermi surface. Figure.\ref{Avsw}(d) is plotted at the boundary of the Brillouin zone on the Fermi surface. 

%The structure of the plots in all the cases are quite similar qualitatively. It involves two high peaks, whose position depends on the momentum components, $(k_x, k_y)$. The peak coincides with $\omega=0$ at the Fermi momentum. These pairs of peaks are accompanied by a velley and a hump like structure on its right hand side. As the Fermi momentum moves away from the Fermi surface, the valley gets flatter and almost vanishes for zero momentum. The gap between the pair of the peaks also varies with the variation of the momentum. 
%
\section{The worldvolume fermion}

In the last two sections we have analysed the fermionic responses to an inhomogenous solution of charge density wave obtained from a D2-D8 system. The inhomogeneous solution has a periodicity and using a generic fermion in that background we find that it exhibits the Fermi surface near the boundary of the Brillouin zone. In addition, as mentioned in the last section, the density plot also shows a circle near the origin of the $k$-plane deep inside the Brillouin zone, which with large enough charge may lead to another Fermi surface.

However, in order to address the response of the fermionic operators in the dual (2+1) dimensional SYM theory in the present set up, it is necessary that, instead of considering the generic fermions, we consider the fermionic fields that arises in the worldvolume theories of the D8 brane. That will be our concern in the present section and we will be using the same methodology that we have described for the generic fermions. The worldvolume  fermionic fields couple to the fermionic opertors living in the dual  gauge theory and by studying the behaviour of these fermionic fields we can explore the dynamics of the fermionic operators.

We will begin with the action of the worldvolume fermions of the D8 brane. The general expression for the action for the worldvolume fermions in $(p+1)$-dimensions has been discussed in \cite{Martucci:2005rb,Marolf:2003vf,Marolf:2003ye}. The action is given by
\be
S_F = \frac{T_p}{2} \int d^{p+1}\xi e^{-\phi} \sqrt{-\det(g+ {\mathcal F})}\bar{\Psi} (1 - \Gamma_{D_p}) [ (\tilde{M}^{-1})^{\alpha\beta} \Gamma_\beta D_\alpha - \triangle ] \Psi
\ee
where ${\mathcal F} = dA$ represents the worldvolume field strength, $T_p^{-1} = (2\pi)^p (\alpha^\prime)^{\frac{p+1}{2}} g_s$ is the brane tension, $\Gamma_\alpha$ is the pullback of the spacetime gamma matrices $\Gamma_m$, given by $\Gamma_\alpha = \Gamma_{\bar{m}} e_{m}^{~\bar{m}} \partial_\alpha x^m$ and for our purpose $\Psi$ is a ten dimensional Majorana spinor in type IIA theory. 

The matrix $\tilde{M}_{\alpha\beta}$  introduces coupling to the metric tensor and the electromagnetic field strength and for the type IIA theory is given by
\be \tilde{M}_{\alpha\beta} = g_{\alpha\beta} + \Gamma_{11} {\mathcal F}_{\alpha\beta} \ee
where $g_{\alpha\beta}$ represents the pull back of the spacetime metric on the worldvolume. The general expressions for $D_m$ and $\triangle$ are discussed in \cite{Martucci:2005rb}. In the present case, we do not have any three-form field strength or two-form field strength in the background. We only need to consider the dilation and the four-form field strength in the background and so the expressions simplify into the following.
\be
\begin{split}
D_m &= \nabla_m  -\frac{1}{8.4!} F_{npqr}\Gamma^{npqr} \Gamma_m ,\\
\triangle &= \frac{1}{2} \Gamma^m\partial_m\phi - \frac{1}{8.4!} e^\phi F_{mnpq}\Gamma^{mnpq}\\
\nabla_m &= \partial_m - \frac{i}{4}( \omega_m)_{\bar{ab}}\Gamma^{\bar{ab}}
\end{split}
\ee
This gives rise to the Dirac equation
\be \label{dirac}
 [ (\tilde{M}^{-1})^{\alpha\beta} \Gamma_\beta D_\alpha - \triangle ] \Psi = 0.
\ee

In order to analsye the Dirac equation we will segregate the coordinates in the following manner.
The D8 brane is wrapped on a product of four dimensional space $(t,x,y,r)$, an $S^2$ and an $S^3$ and it is transverse to the direction $\psi$. We will use $x^\mu$, $\zeta^i$ and $\eta^a$ to denote coordinates along the four dimensional space, the $S^2$ and the $S^3$ respectively.
\be\begin{split}
x^\mu &= t, x, y, r,\quad \mu=0,1,2,3\\ 
& \zeta^i = (\theta_1, \phi_1) ,\quad i=1,2 \\
& \eta^a = (\xi, \theta_2, \phi_2) ,\quad a = 1,2,3.
\end{split}
\ee

We choose the ten dimensional gamma matrices in the following manner.
\be\begin{split} \label{gamma}
\Gamma^{\bar\mu} &= \sigma_2 \otimes 1_2 \otimes 1_2 \otimes \gamma^{\bar{\mu}},\quad \bar{\mu}=0,1,2,3\\
\Gamma^{\bar\psi} &= \sigma_2 \otimes 1_2 \otimes 1_2  \otimes \gamma_5\\
\Gamma^{\bar i} &= \sigma_1 \otimes \sigma_i \otimes  1_2 \otimes 1_4\quad \bar{i}=1,2 \\
\Gamma^{\bar a} &= \sigma_1 \otimes \sigma_3 \otimes  \sigma_a \otimes 1_4 \\
\Gamma_{11} &= \sigma_3 \otimes 1_2 \otimes  1_2 \otimes 1_4, \quad \bar{a} = 1,2,3.
\end{split}\ee
where the ${\bar\mu}, {\bar\psi} , {\bar i}$ and ${\bar a}$ represents the respective tangent space directions.

The ten dimensional background metric is given in \ref{metric} and according to that the Dirac equation (\ref{dirac}) can be splitted as
\be\begin{split}
&\Big( \big[ (\tilde{M}^{-1})^{\mu\nu} \Gamma_\nu D_\mu + (\tilde{M}^{-1})^{ij} \Gamma_j D_i + (\tilde{M}^{-1})^{ab} \Gamma_b D_a \big] 
- \frac{e^\phi}{8} \big[ (\tilde{M}^{-1})^{\mu\nu} \Gamma_\nu \not{F_4} \Gamma_\mu 
\\
&+ (\tilde{M}^{-1})^{ij} \Gamma_j \not F_4 \Gamma_i + (\tilde{M}^{-1})^{ab} \Gamma_b \not F_4 \Gamma_a - \not F_4 \big] - \frac{1}{2} \Gamma^m\partial_m \phi \Big) \Psi = 0
\end{split}\ee
where we have used $\not{F}_4 = \frac{1}{4!} \Gamma^{npqr}F_{npqr}$. 
For the present background solution the totally antisymmetric four-form field strength can be written as
\be \label{four-form} (F_4)_{\bar{t}\bar{x}\bar{y}\bar{u}} = -5 \frac{r_T^{3/2}}{u^{3/2}}. \ee

The contributons arising from the two-sphere and the three-sphere can be further simplified and are written as follows. Since on the three-sphere there is no worldvolume flux, we have 
\be\begin{split}
 (\tilde{M}^{-1})^{ab} \Gamma_b D_a &= \frac{r_T^{1/4}}{u^{1/4}\cos\psi} \sigma_1\otimes\sigma_3\otimes\not{D}_{S^3}\otimes 1_4, \\
 (\tilde{M}^{-1})^{ab} \Gamma_b \not F_4 \Gamma_a &= 3 \not F_4,
 \end{split} \ee
 where $\not{D}_{S^3}$ is the Dirac operator on the $S^3$ with the round metric.
 
 On the two-sphere there is a magnetic field as given in (\ref{magneticfield}) introduced for the stabilty and we obtain
 \be\begin{split}
 (\tilde{M}^{-1})^{ij} \Gamma_j D_i &= \frac{r_T^{1/4}}{u^{1/4} \sin\psi} \frac{(1 + ip\sigma_3\otimes\sigma_3\otimes 1_2\otimes 1_4)}{1+p^2} \sigma_1\otimes\not{D}_{S^2}\otimes 1_2 \otimes 1_4, \\
 (\tilde{M}^{-1})^{ij} \Gamma_j \not F_4 \Gamma_i &= 2 \frac{(1 + ip\sigma_3\otimes\sigma_3\otimes 1_2\otimes 1_4)}{1+p^2} \not F_4,
\end{split} \ee
where in order to make the expressions compact we have introduced $p$, given by,
\be
p = \frac{\hat{b}}{2\pi\alpha^\prime u^{1/2} \sin^2\psi},
\ee
and $\not{D}_{S^2}$ is the Dirac operator on the $S^2$ with the round metric.
Substituting the expression of the four-form field strength (\ref{four-form}) and the gamma matrices (\ref{gamma}) we obtain
\be \not F_4 = i \kappa 1_2\otimes 1_2 \otimes 1_2 \otimes \gamma_5 ,\quad \kappa = \frac{5 r_T^{3/2}}{u^{3/2}}.
\ee

 Substituting all these expressions the Dirac equation reduces to
 
\be\begin{split}
& (g^\mu_{~\lambda} - \F^\mu_{~\lambda} \sigma_3 \otimes 1_2 \otimes  1_2 \otimes 1_4 ) [ (g - \F.g^{-1}.\F)^{-1}]^{\lambda\nu} \sigma_2\otimes 1_2 \otimes  1_2 \otimes \gamma_\nu D_\mu 
\\
& + \frac{r_T^{1/4}}{u^{1/4}} [ \frac{(1 + i p \sigma_3 \otimes \sigma_3 \otimes  1_2 \otimes 1_4)}{\sin\psi (1+ p^2)}\sigma_1\otimes\not{D}_{S^2}\otimes 1_2 \otimes 1_4
+\frac{1}{\cos\psi} \sigma_1\otimes\sigma_3\otimes\not{D}_{S^3}\otimes 1_4]
\\
& - i \frac{\kappa}{8} e^\phi [ - (g^\mu_{~\lambda} - \F^\mu_{~\lambda} 1_2 \otimes 1_2 \otimes  1_2 \otimes \gamma_\nu\gamma_\mu\gamma_5 ) [ (g - \F.g^{-1}.\F)^{-1}]^{\lambda\nu} 1_2\otimes 1_2 \otimes  1_2 \otimes \gamma_\nu \gamma_\mu 
\\
&+ 2 \frac{(1 + ip\sigma_3\otimes\sigma_3\otimes 1_2\otimes 1_4)}{1+p^2} )  1_2 \otimes 1_2 \otimes  1_2 \otimes \gamma_\nu\gamma_\mu\gamma_5]\Psi = 0,
\end{split}\ee
 
We introduce the two component spinor for the first $2\times 2$ matrices in the expressions of the gamma matrices as
\be
\lambda^0 = \left(\begin{array}{c} 1\\ 0 \end{array}\right)
\quad
\lambda^1 = \left(\begin{array}{c} 0\\ 1 \end{array}\right)
\ee
which are eigenvectors of $\sigma_3$. 

Since the Dirac operator involves ${\not{D}}_{S^3}$ and ${\not{D}}_{S^2}$, it will be convenient to expand the spinor in terms of the eigenfunctions of those operators. Eigenfunctions of the Dirac operators on general spheres have been discussed in \cite{Camporesi:1995fb} and for our purpose we only need the eigenvalues for $S^2$ and $S^3$. For our convenience we choose a slightly different notation and for the two-sphere we write
\be\begin{split}
{\not{D}}_{S^2} \xi^{(s)}_l &= i (-1)^s (l+1)  \xi^{(s)}_l ,\quad s=0,1,\quad l= 0,1,2,...,
\\ 
\sigma_3  \xi^{(s)}_l &= (-1)^s \xi^{(1-s)}_l.
\end{split}
\ee
Similarly for the Dirac spinors on the three-sphere we write
\be
{\not{D}}_{S^3} \eta^{(t)}_m = i (-1)^t (m+3/2)  \eta^{(t)}_m , \quad t=0,1, \quad m= 0,1,2,...
\ee

We expand the ten dimensional spinor as
\be
\Psi = \lambda^\alpha \otimes \xi^s_l \otimes \eta^t_m \otimes \chi[\alpha, s, t, l,m]
\ee
where in this notation, $\alpha, s, t = 0, 1$, while $l,m = 0,1,2,...$ and each $\chi[\alpha, s, t, l,m]$ represents a four-dimensional spinor. We substitute this expression in the Dirac equation and setting the coefficient of each $ \lambda^\alpha \otimes \xi^s_l \otimes \eta^t_m$ equal to zero we obtain the following equation for $\chi^{\alpha, s, t}_{l,m}$.
\be
\begin{split} \label{eqn}
& i N_1^{\mu\bar{\nu}} \gamma_{\bar{\nu}} D_\mu \chi [1-\alpha,s,t,l,m] \\
&+ i \frac{r_T^{1/4}}{u^{1/4}} \Big\lbrace~ \frac{(-1)^s (l+1)}{\sin\psi (1 + p^2)} \Big( \chi[1-\alpha,s,t,l,m] + i (-1)^\alpha p \chi[1-\alpha,1-s,t,l,m] \Big)\\
& + \frac{(-1)^t (m + 3/2)}{\cos\psi} \chi[1-\alpha, 1-s, t, l,m] ~ \Big\rbrace\\
&-i \frac{5 g_s r_T^{1/4}}{8u^{1/4}} \Big\lbrace ~ - N_2^{\bar{\mu}\bar{\nu}}\gamma_{\bar{\nu}} \gamma_{\bar{\mu}}\gamma_5 \chi[\alpha,s,t,l,m] + 2 \Big(1 + \frac{1}{1+p^2} \Big)\gamma_5\chi[\alpha,s,t,l,m] \\
&+ \frac{2ip}{1+p^2}(-1)^\alpha \gamma_5 \chi[\alpha,1-s,t,l,m] ~ \Big\rbrace\\
&-i \frac{5 r_T^{1/4}}{8 u^{1/4}} \sqrt{f} \Big[ 1 + \frac{u^3 \psi_x^2}{1+u^3 \psi_x^2 + u^2 f \psi_u^2}\Big]^{1/2} (-1)^\alpha \gamma^{\bar{u}} \chi[1-\alpha,s,t,l,m] = 0
\end{split}
\ee
where, in the above equation we have used
\be\begin{split}
N_1^{\mu\bar{\nu}} &= (g^{-1})^{\mu\kappa} [ (-1)^\alpha g + {\mathcal F} ]_{\kappa\lambda} [(g- {\mathcal F}.g^{-1}.{\mathcal F})^{-1}]^{\lambda\nu} e_\nu^{~\bar{\nu}},\\
N_2^{\mu\bar{\nu}}& = e^{\bar{\mu}}_{~\mu} (g^{-1})^{\mu\kappa} [ g - (-1)^\alpha {\mathcal F} ]_{\kappa\lambda} [(g- {\mathcal F}.g^{-1}.{\mathcal F})^{-1}]^{\lambda\nu} e_\nu^{~\bar{\nu}},
\end{split}\ee
In order to obtain the spectral density of the fermionic operators living on the boundary field theory we need to to solve these equations with appropriate boundary conditions and study the asymptotic behaviours. In terms of $u$ the horizon is located at $u=1$ and the boundary is located at $u=0$.

The momentum mode expansion of the different spinor components  $\chi[1-\alpha,s,t,l,m]$ are given by
\be
\chi[\alpha,s,t,l,m]  = \sum\limits_{n \in Z} \chi[\alpha,s,t,l,m]^{(n)}  e^{ i n K x }.
\ee 
In the following discussion, we will keep the momentum modes implicit.

$\bullet${\bf Near horizon behaviour:} The necessary boundary conditions follow from the near horizon behaviour. Taking the near horizon limit $u\rightarrow 1$ of the above equations and keeping only the leading order terms, we obtain
\be 
\Big( \pa_u - \frac{i\omega}{5(1-u)} \gamma_{\bar{t}} \gamma_{\bar{u}} \Big) \chi[\alpha,s,t,l,m] = 0,
\ee
For the four domensional gamma matrices $\gamma^\mu$ we have adopted the following choice \cite{cremonini4} as given below in $2\times2$ block form using the Pauli spin matrices . 
\be\begin{split}
\gamma^{\bar{t}} &=\left(\begin{array}{cc}i\sigma_1&0\\0&i\sigma_1\end{array}\right),\quad
\gamma^{\bar{u}}  = \left(\begin{array}{cc} - \sigma_3&0\\0&-\sigma_3\end{array}\right), \quad
\gamma^{\bar{x}} = \left(\begin{array}{cc} - \sigma_2&0\\0&\sigma_2\end{array}\right), \\
\gamma^{\bar{y}} &=  \left(\begin{array}{cc} 0&  \sigma_2\\ \sigma_2&0 \end{array}\right) , \quad
\gamma_5 = i \gamma^{\bar{t}} \gamma^{\bar{u}}\gamma^{\bar{x}}\gamma^{\bar{y}}= \left(\begin{array}{cc} 0& i \sigma_2\\ -i \sigma_2&0 \end{array}\right),
\end{split}\ee
 After substituting the expressions for the background metric and gauge fields and choosing the near horizon limit we obtain,
\be
\begin{split}
\chi[\alpha,s,t,l,m]_1 & \sim A[1,\alpha,s,t,l,m] (1-u)^{\pm \frac{i\omega}{5}} ,\quad \chi[\alpha,s,t,l,m]_2 \sim B[1,\alpha,s,t,l,m] (1-u)^{\pm \frac{i\omega}{5}} \\
\chi[\alpha,s,t,l,m]_3 & \sim A[2,\alpha,s,t,l,m] (1-u)^{\pm \frac{i\omega}{5}} ,\quad \chi[\alpha,s,t,l,m]_4 \sim B[2,\alpha,s,t,l,m] (1-u)^{\pm \frac{i\omega}{5}}
\end{split}
\ee
We have chosen the minus sign and impose the in-falling boundary conditions \cite{Iqbal:2009fd} as that is the correct choice for holographic computation of retarded Green's function of the dual theory living at the boundary.  Furthermore, the equations at the near horizon limit also implies
\be 
B[r,\alpha,s,t,l,m]  = - i A[r,\alpha,s,t,l,m] .\label{bc3}
\ee

$\bullet${\bf Asymptotic behaviour:}
We can obtain the asymptotic behaviour of the fermions by considering the $u\rightarrow 0$ limit of the equations that we obtain above in (\ref{eqn}) and keeping only the leading order terms. In this limit the equations get simplified considerably and assumes the following form,
%\be\begin{split} i (-1)^\alpha u^{3/4} \nabla_u \gamma^{\bar{u}} \chi[1-\alpha,s,t,l,m] + i (-1)^t \frac{m+3/2}{\cos\psi_0} u^{-1/4} \chi[1-\alpha,1-s,t,l,m] + i \frac{5g_s}{4 u^{1/4}} \chi[\alpha,s,t,l,m]  -i (-1)^\alpha \frac{5}{8 u^{1/4}} \gamma^{\bar{u}} \chi[1-\alpha,s,t,l,m] =0 \end{split}\ee

\be\begin{split} 
i (-1)^\alpha  \big( u \nabla_u - \frac{5}{8} \big) \gamma^{\bar{u}} \chi[1-\alpha,s,t,l,m] &+ i (-1)^t (m+3/2)  \chi[1-\alpha,1-s,t,l,m] 
\\ &+ i \frac{5g_s}{4} \gamma_5 \chi[\alpha,s,t,l,m]  = 0.
 \end{split}\ee
 
 These equations are first order coupled differential equations and can be disentangled by making the following linear combinations of verious components of the spionors.
 \be\begin{split}
 \eta_A [\alpha, t, a, b] &= \frac{1}{4} [ (\chi_1[1-\alpha, s, t] + (-1)^a \chi_1[1-\alpha, 1-s,t] ) \\
 & + (-1)^b  (\chi_4[\alpha,s,t] + (-1)^a \chi_4[\alpha, 1-s, t] ) ] , \\
  \eta_B [\alpha, t, a, b] &= \frac{1}{4} [ (\chi_3[1-\alpha, s, t] + (-1)^a \chi_3[1-\alpha, 1-s,t] ) \\ 
  &+ (-1)^b  (\chi_2[\alpha,s,t] + (-1)^a \chi_2[\alpha, 1-s, t] ) ].
 \end{split}\ee
 The behaviour of the functions $\eta_A$ and $\eta_B$ are given by
 
  In terms of the new functions $\eta_A[\alpha,a,b,t]$, $\eta_B[\alpha,a,b,t]$ the asymptotic behaviour can be written as
  \be\begin{split}
   \eta_A[\alpha,t,a,b] &\sim u^{ [(-1)^{(t+\alpha+a)}(m+\frac{3}{2}) + (-1)^{b + \alpha} \frac{5 g_s}{4} + \frac{5}{8}]}\\
    \eta_B[\alpha,t,a,b] &\sim u^{[(-1)^{(t+\alpha+a)}(m+\frac{3}{2}) - (-1)^{b + \alpha} \frac{5 g_s}{4} + \frac{5}{8}]}
   \end{split}\ee
   Expressing the spinor components at the asymptotic limits in terms of linear combinations of  $\eta_A[\alpha, t, ,a,b]$ and $\eta_B[\alpha,t,a,b]$  we can identify the asymptotic behaviours of the spinor components. In contrast to the earlier case, where the asymptotic limits of the spinor components are characterised with two different powers of $u$, here it will involve four different powers of $u$ in general. 
     
 For the convenience of the numerical computations, we will further introduce the notation that
 \be\begin{split}
 \chi_1[\alpha,s,t] = \chi_1[1,\alpha,s,t],\quad  \chi_3[\alpha,s,t] = \chi_1[2,\alpha,s,t] \\
  \chi_2[\alpha,s,t] = \chi_2[1,\alpha,s,t],\quad  \chi_4[\alpha,s,t] = \chi_2[2,\alpha,s,t] 
  \end{split}\ee
  
We will follow the same method adopted in the last section to compute the Green's function. Here the boundary conditions can be denoted by $(r, \alpha, s, t)$, which corresponds to the in-falling boundary condition on the spinor component $\chi_1[r,\alpha,s,t]$ and  $\chi_2[r,\alpha,s,t]$ and setting all other spinor components to be zero. We will write the coefficient of the having highest negative (positive) power of $u$ in the asymptotic expression of $\chi_2[r,\alpha,s,t]$ ($\chi_1[r,\alpha,s,t]$) as $b[r,\alpha,s,t] $ ($a[r,\alpha,s,t] $) with $(r',\alpha' , s',t')$ boundary condition as \\
$b (r,\alpha,s,t ; r',\alpha' , s',t' )$ ($a(r,\alpha,s,t ; r',\alpha' , s',t' )$) and in this notation the expression for Green's function (\ref{green1}) can be written as 
 \be\begin{split}
 a(r,\alpha,s,t ; r',\alpha' , s',t' ) =
 \\
 & \sum\limits_{r'',\alpha'' , s'',t''} G^R(r,\alpha,s,t ;r'',\alpha'' , s'',t'' )  b(r'',\alpha'' , s'',t'' ; r',\alpha' , s',t').  \label{green1}
 \end{split}
 \ee
 Writing $a(r,\alpha,s,t ; r',\alpha' , s',t' )$ and $b(r'',\alpha'' , s'',t'' ; r',\alpha' , s',t')$ as matrices {\bf a} and {\bf b}, the relation becomes
 \be
 {\bf a} = G^R . {\bf b}, \quad\quad   G^R = {\bf a}.{\bf b}^{-1}
 \ee
 As in the section 3, we will assume that the diagonal momentum mode dominates and furthermore we will keep only the $n=0$ modes. Then the spectral functions follows from the Green's function $G^R$.
 
\begin{figure}[h]
			\centering
			%\begin{subfigure}{8cm}
				%\centering
				\includegraphics[width=8cm]{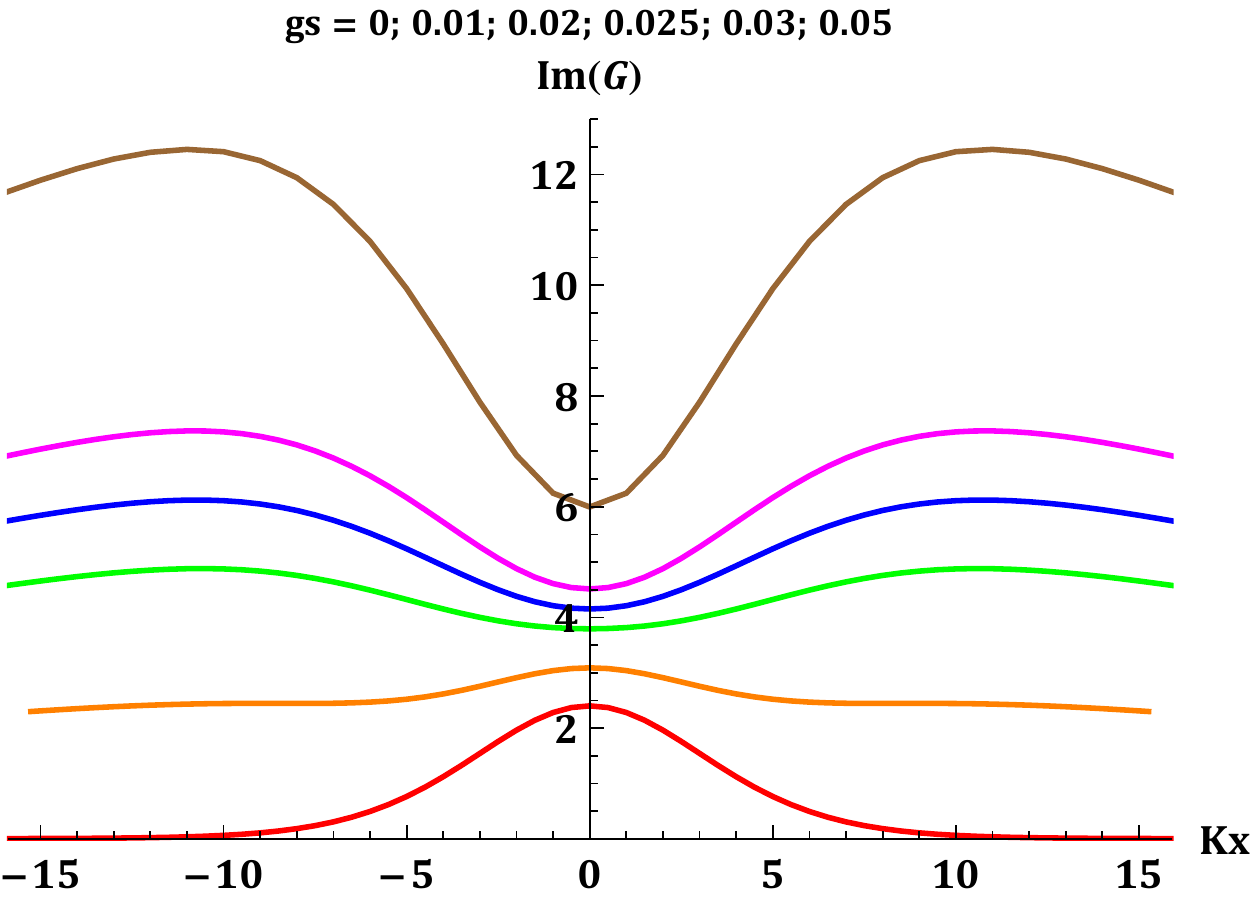}
						%\caption{}
				%\label{fig:1a}
			%\end{subfigure}%
			\caption{Plot of spectral function A vs. $k_x$ with $k_y=0$ for worldvolume fermions. From bottom to top: $g_s=$ 0 (red), 0.01 (orange), 0.02 (green), .025 (blue), .03 (magenta), .05 (brown)}
			\label{FAvkx}
		\end{figure}

 We have solved the Dirac equations numerically and plotted the spectral functions obtained in the present case in Fig.\ref{FAvkx} for several values of $g_s$ keeping $l=m=0$. As one can observe, for small value of $g_s$, it will be similar to  $g_s=0$ and it shows a maximum around $k_x=0$ and for the entire range the height is quite small. However, the peak is quite flat and it does not qualify as a Fermi surface. As $ g_s$ increases, the maxima at $k_x=0$ disappears and a pair of maxima develop located symmetrically on both sides of $k_x=0$. The overall value of the spectral function also increases with increase of $g_s$. 
 
\begin{figure}[h]
			\centering
			\begin{subfigure}{8cm}
				\centering
				\includegraphics[width=7cm]{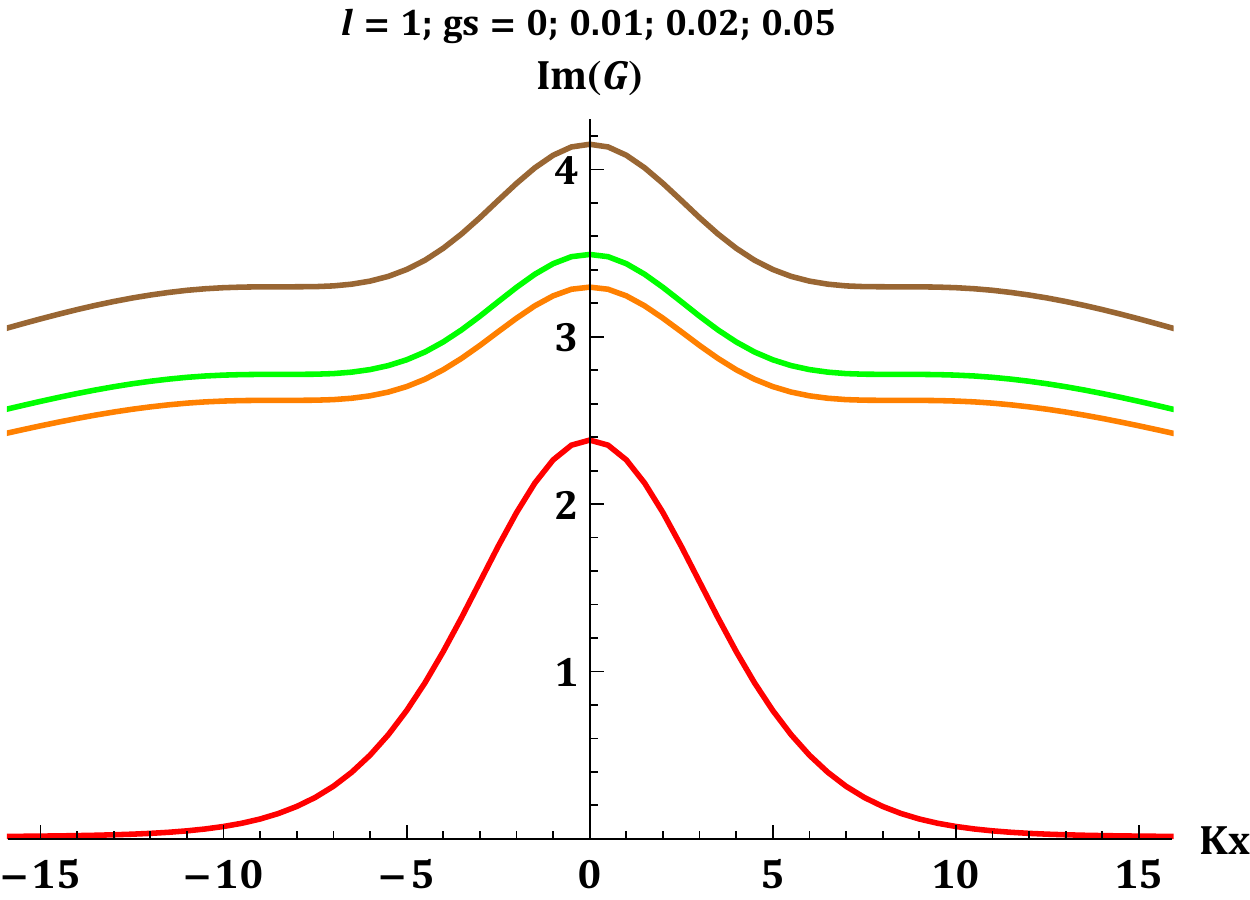}
						\caption{}
				%\label{fig:1a}
			\end{subfigure}%
			\begin{subfigure}{8cm}
				\centering
				\includegraphics[width=7cm]{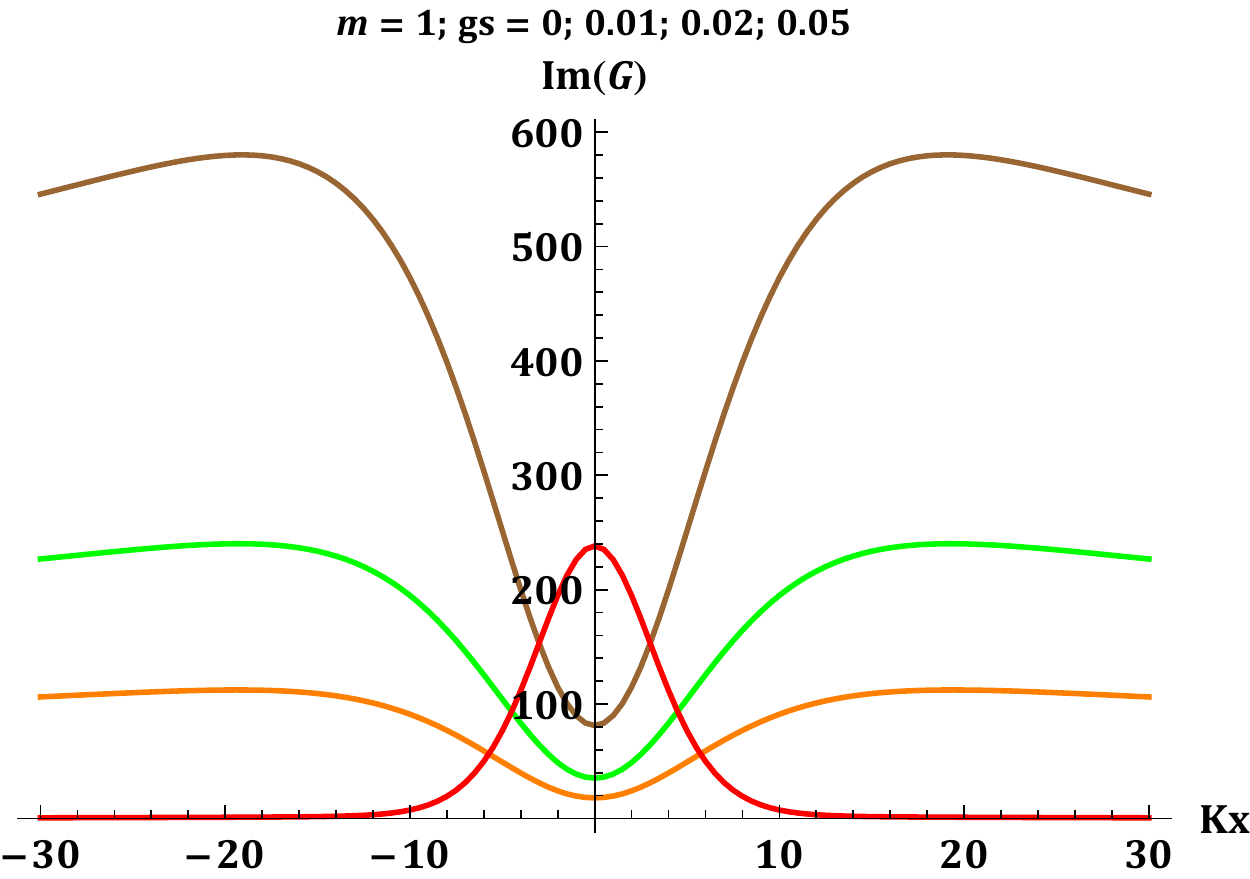}
						\caption{}
			%	\label{fig:1b}
			\end{subfigure}
			\caption{Plot of spectral density A vs. $w$ with $g_s=$ 0 (red), 0.01 (orange), 0.02 (green), .05 (brown) (a) $ l=1, m=0$, (b) $l=0, m=1$}
			\label{FAvlm}
		\end{figure}

The eigenvalues of the Dirac operators along $S^2$ and $S^3$ introduces running mass-like terms in the Dirac equation. In the case of usual Dirac operators, as we have seen in the last section where generic fermions were discussed, the asymptotic behaviour of the Dirac equation is determined by the mass terms. In the present case, however, though the eigenvalues of the Dirac operator along $S^3$ contributes to the the asymptotic limit of the Dirac equation, the one associated with the $S^2$ does not contribute. We have plotted the spectral function for severl values of $g_s$ in Fig. \ref{FAvlm} for (a)  $l=1$, $m=0$ (a)  $l=0$, $m=1$. For $l=1$ it shows a maximum around $k_x=0$ and qualitatively it remains the same for other values of $g_s$. For $m=1$ once again it shows a maximum at $k_x=0$ for small $g_s$, which turns into a local mimimum with increase of $g_s$. Though there are maxima, but they are too broad to be qualified as a Fermi surface. 

Since the worldvolume fermions of the D8-brane couple to the worldvolume gauge fields in adjoint representation and since we have  a U(1) gauge field the worldvolume fermions are not charged with respect to the gauge fields. As we have seen in the discussion of generic fermion for zero charge we cannot expect a Fermi surface. However, there we observed small and broad peak around the origin and as we mentioned over there, for adequate value of the charge it may lead to a Fermi surface. Similarly one can expect that once the fermions have adequate charge, it may lead to a Fermi surface. Since increasing $g_s$ can be considered as including the effect of $\frac{1}{N}$ terms in the field theory living on the boundary, if the same behaviour persists,  it indicates that the $\frac{1}{N}$ effects will modify the structure of the Fermi surface. In order to study it we need to consider at least two D8 branes, so that it can lead to a gauge group $SU(2)$ to which the worldvolume fermions couple with non-zero charge.

\section{Discussion}

In this work we have considered fermionic response of a spatially modulated solution obtained in a top-down approach. Intersecting D2-D8 brane system develops instability and for a particular region of the parameter space leads to a solution which consists of a charge density wave and a spin density wave. We set the chemical potential to be inside the region of stability and consider that gravitational solution as the background. In this background we introduce fermions and numerically solve the Dirac equations. From the asymptotic behaviour of the solutions, we obtain the spectral density associated with the fermionic operators in the dual theory living on the boundary. 

We begin with introducing generic fermions assuming they will couple to the world volume gauge field, keeping the charges flexible.  In order to study the Fermi surfaces we look for appropriate peak of spectral density function. Due to spontaneous breaking of the translational symmetry, the background solutions are characterised with a periodicity determined by the minimum of the free energy, which in turn depsends on the chemical potential. On top of this solution we have also considered simulating an ionic lattice by introducing periodicity in the chemical potential by hand, which leads to an explicit breaking of translation symmetry. We find as the fermionic charge increases, the height of the peak increases. It shows once the fermionic charge is large enough the Fermi surface materializes. In the present model, plotting the spectral density over momentum plane we obtain the Fermi surface to be a series of circles distributed over the Brillouin zones. As the charge increases further, the circular surfaces get bigger, cross the boundary of the Brillouin zone and intersect. We find at the points of the intersections, they develop gap leading to the Fermi pockets (inner Fermi surface). In presence of the ionic lattice, the gap becomes more pronounced leading to a wider separation. The height of the spectral density plot is also keep on diminishing with the increasing strength of the ionic lattice. This indicates that for a larger value of the strength of the ionic lattice, the Fermi pockets will disappear.
Similar study of Fermi surfaces for a charge density model has appeared in the bottom-up approach \cite{cremonini4}. They also obtained similar behaviour of Fermi surfaces suggesting these are generic rather than model dependent features of the charge density wave solutions. 

Next we consider the worldvolume fermions in the same background. Since worldvolume fermions transform under the adjoint representation, charge of the usual couping with the gauge field is zero in this case. Nevertheless, due the structure of the action, the electromagnetic flux enters the Dirac equation. We studied the plot of spectral function and find it shows a maximum around the origin. We also observe that with the increase of $g_s$ the maximum of spectral function, separates into a pair of maxima, moving away from the $k_x=0$. If we consider that the increasing values of $g_s$ would correspond to $1/N$ effects in the dual field theory this may hint at the possible effects of $1/N$ correction. However, the maximum does not qualify as a Fermi surface, which is due to the fact that the charge of the fermions coupling to the gauge field is zero. Nevertheless, one can expect a similar behaviour will persist for adequate charge. 

A natural extension of the present work is to consider the worldvolume fermions in the D2-D8 intersecting brane model with a gauge group of higher rank, such as $SU(2)$ and examine the behaviours of the fermions as has been done in \cite{Ammon:2010pg} in a linearised approximation. Being a top-down approach it is possible to explicitly consider the exact field theory model dual to it living on the boundary. It will be quite interesting to understand these features in a field theory set-up in the dual model. It will shed light on the formation of the Fermi surfaces  from the perspective of a field theoretic understanding. The behaviour of the spectral density function with the variation of the frequency also merits a field theoretic study. 
%In fact, a similar study of the spectral functions of the  CDW phonons in the case of a CDW superconductor \cite{lei}  shows quite a similar structure. 

We can also extend the analysis for the Majorana fermions. Supergravity models often introduce Pauli coupling, which in holographic models introduces gaps in the Fermi surface. It may be interesting to study the implications of the Pauli couplings for the generic fermions in the present model and how it modifies the Fermi surface. Lastly, it may be mentioned that
the spatially modulated solution that we have considered here is obtained in a probe approximation. One can obtain the full solution by considering the gravitational back reaction, which may provide a more complete picture of the Fermi surface in the present context. 
%%%%%%%%%%%%%%%%%%%%%%%%%
%
%%%%%%%%%%%%%%%%%%%%%%%%%%%
			%%%%%%%%%%%%%%%%%%%%%%%%%%%%%%%%%%%%%%%%%%%%%%
			\section*{Acknowledgement}
SM thankfully acknowledges the financial assistance received from Science and Engineering Research Board (SERB), India (project file no. CRG/2019/002167).
\newpage

			%%%
%\bibliographystyle{ieeeter}
\bibliographystyle{plain}

		\end{document}